\pdfoutput=1
\documentclass[11pt,twoside,a4paper,cmspaper,final,collab]{cms-tdr}
\def\svnVersion{8f1cf76}\def\svnDate{2025/04/10}\def\cmsCernNoTag{CERN-EP-2024-273}\def\cmsCernDate{\today}\def\cmsMessage{Published in the Journal of Instrumentation as \href{http://dx.doi.org/10.1088/1748-0221/20/04/P04012}{\doi{10.1088/1748-0221/20/04/P04012}.}}
\begin{document}\cmsNoteHeader{SMP-23-006}

\renewcommand{\cmsCollabName}{The CMS and TOTEM Collaborations}
\renewcommand{\cmsTag}{CMS-\cmsNUMBER\\&TOTEM 2024-002\\}
\renewcommand{\cmsPubBlock}{\begin{tabular}[t]{@{}r@{}l}&CMS \cmsSTYLE~\cmsNUMBER\\&TOTEM 2024-002\\\end{tabular}}
\renewcommand{\cmsCopyright}{\copyright\,\the\year\ CERN for the benefit of the CMS and TOTEM Collaborations.}
\renewcommand{\appMsg}{See Appendices~\ref{app:collab} and~\ref{app:totem} for the lists of collaboration members}
\renewcommand{\cmslogo}{\includegraphics[height=2cm]{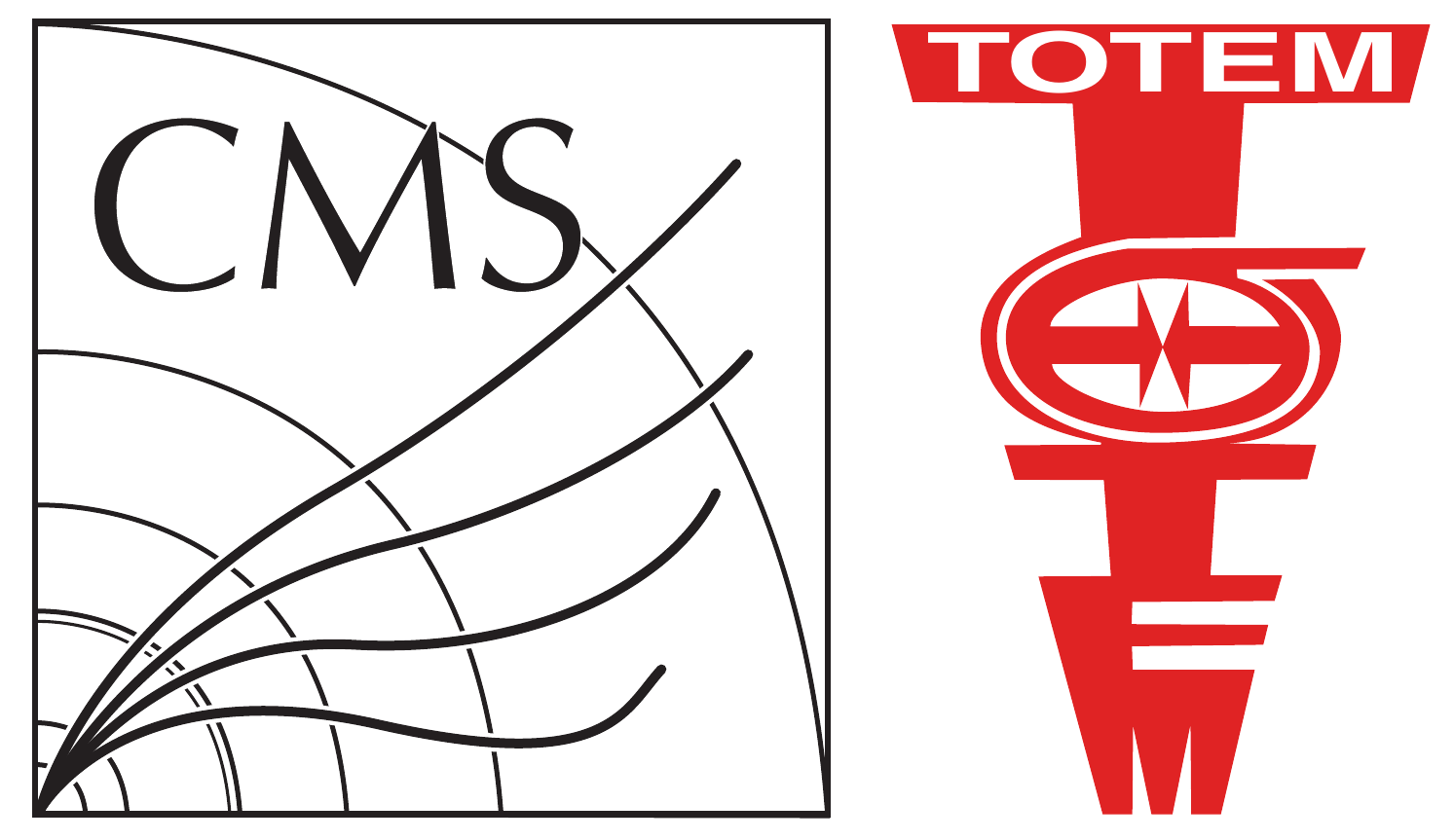}}

\cmsNoteHeader{SMP-23-006}
\title{Proton reconstruction with the TOTEM Roman pot detectors for high-$\beta^*$ LHC data}

\date{\today}

\abstract{The TOTEM Roman pot detectors are used to reconstruct the transverse momentum of scattered protons and to estimate the transverse location of the primary interaction. This paper presents new methods of track reconstruction, measurements of strip-level detection efficiencies, cross-checks of the LHC beam optics, and detector alignment techniques, along with their application in the selection of signal collision events. The track reconstruction is performed by exploiting hit cluster information through a novel method using a common polygonal area in the intercept-slope plane. The technique is applied in the relative alignment of detector layers with $\mum$ precision. A tag-and-probe method is used to extract strip-level detection efficiencies. The alignment of the Roman pot system is performed through time-dependent adjustments, resulting in a position accuracy of 3\mum in the horizontal and 60\mum in the vertical directions. The goal is to provide an optimal reconstruction tool for central exclusive physics analyses based on the high-$\beta^*$ data-taking period at $\sqrt{s}=13\TeV$ in 2018.}

\hypersetup{%
pdfauthor={CMS Collaboration},%
pdftitle={Proton reconstruction with the TOTEM Roman pot detectors for high-beta LHC data},%
pdfsubject={TOTEM},%
pdfkeywords={CMS, TOTEM, proton reconstruction}}

\maketitle 

\section{Introduction}
\label{sec:intro}

Elastic and diffractive interactions of high-energy protons at the CERN LHC are
characterised by scattered protons emerging at small angles (microradians) with
respect to the beam direction. These protons can be measured with detectors
that are inserted into the beam pipe called ``Roman pots'' (RPs), which are
movable near-beam devices that allow the detectors to be brought very close
(down to a few \mm) to the beam without affecting the vacuum, beam stability, or
other aspects of accelerator operations.

This paper describes methods for the reconstruction of the scattered proton
trajectories in such near-beam detectors. These reconstruction techniques have
been developed by the CMS and TOTEM Collaborations for the studies of central
exclusive hadron
production~\cite{TheTOTEMCollaboration:2014nkc,Osterberg:2014mta,TOTEM:2024dzz},
where in addition to the two scattered protons, a central system of particles
is created. This paper is based on data jointly taken by the CMS and TOTEM
detectors in a special, low instantaneous luminosity run in July 2018, at the
proton-proton (pp) centre-of-mass energy of 13\TeV, with an integrated
luminosity of $4.7\unit{pb}^{-1}$~\cite{CMS:2019jhq}.

The acceptance of the RP detector system for elastically or diffractively
scattered protons is determined by the beam optics configuration.

With the high-$\beta^*$ setting, the beam divergence is small. Here
$\beta^*$ denotes the value of the amplitude function at the interaction
point (IP), often referred to as the distance from the IP where the beam
width doubles~\cite{wiedemann2013particle}. The forward detectors, a
few hundred metres away from the interaction point and within
millimetres from the outgoing beam, are able to measure the scattered
protons at microradian angles and sub-\GeV transverse momenta. The acceptance
as a function of the squared four-momentum transfer of the protons ($t$)
starts at $-0.03\GeV^2$. The proton kinematic properties are determined from
the proton scattering angle, independently of the momentum loss if the latter
is small.  Protons with fractional momentum loss up to 20\% can be observed,
with wide acceptance for the invariant mass of the created central hadron
system~\cite{TheTOTEMCollaboration:2014nkc,Osterberg:2014mta}.

The present paper covers subjects ranging from signal clustering in the silicon
microstrip detectors used in the RPs, to the reconstruction of proton tracks,
the cross-check of the beam optics, and the alignment with respect to the beam
line. The aim is to measure the scattered proton momentum and the location of
the primary interaction for central exclusive physics
events~\cite{Albrow:2010yb}. Improvements relative to earlier methods are
presented for the selection and analysis of signal events, specifically for the
precise momentum balance of forward protons with centrally produced particles.
Several of the discussed techniques are relevant to other beam optics settings
as well.

The paper is organised as follows. An overview of the LHC beam parameters and
the RP detectors is given in Section~\ref{sec:overview}, along with some
considerations on the scattered proton trajectory, and details of the data set
used. The reconstruction of the proton hit locations within the RPs is
discussed in Section~\ref{sec:tracklet}, including pattern recognition and
matching, relative alignment of the detector layers, extraction of the strip
hit efficiencies, and determination of the correction factors. The
reconstruction of the full proton trajectory is detailed next
(Section~\ref{sec:proton}), together with a discussion of the cross-checks of
the beam optics, the distributions of the interaction point locations and
scattering angles, and the absolute alignment of the RPs. Results on momentum
resolution and applications for event classification and selection for physics
analyses are also shown. The paper ends with a summary in
Section~\ref{sec:summary}.

\section{The CMS and TOTEM detectors}
\label{sec:overview}

We use a right-handed coordinate system, with the origin at the nominal
collision point, the $x$ axis pointing to the centre of the LHC ring, the $y$
axis pointing up (perpendicular to the LHC plane), and the $z$ axis along the
anti-clockwise beam direction.

\subsection{The LHC beam parameters}
\label{sec:lhc_beam}

In this low instantaneous luminosity data-taking period, the $\beta^*$ values
in the horizontal and vertical directions are $\beta^*_x =$ 45\unit{m} and
$\beta^*_y =$ 90\unit{m}, respectively. The initial normalised emittance is in the
range $\varepsilon_n = 1.2\text{--}1.6\mum$~\cite{Burkhardt:2019igr}. The
half-crossing angle in the horizontal plane is 60\,$\mu$rad. At the nominal IP,
the transverse size of the beam is $\sigma = \sqrt{\varepsilon\beta^*}$ and the
angular beam size (beam divergence) is $\sigma' = \sqrt{\varepsilon/\beta^*}$,
where the emittance $\varepsilon$ is related to the normalised emittance as
$\varepsilon_n = \varepsilon \beta\gamma$. Here $\beta$ is the ratio of the
velocity of the beam particles to the speed of light, and $\gamma$ is the
Lorentz factor. Given the beam energy of $p_\text{beam} = 6500\GeV$, we have
$\beta\gamma \approx 6930$ and the expected value of the emittance is
$\varepsilon \approx 0.2$\,nm. The transverse momentum spread is
$\sigma_{p_{x/y}} = p_\text{beam} \sigma'_{x/y}$. Table~\ref{tab:beam}
summarises the relevant beam parameters.

The transverse size of the beam spot (BS) can be calculated from the beam width
as $\sigma^\text{BS} = \sigma/\sqrt{2}$. The size of the beam spot is measured
by using reconstructed interaction vertices with tracks in the central silicon
tracker of the CMS detector~\cite{Chatrchyan:2008zzk}. It slowly increases
within an LHC fill from $70$ to $110\mum$ in $x$, and from $90$ to $110\mum$ in
the $y$ direction. With an assumed average initial normalised emittance
$\varepsilon_n \approx 1.4\mum$ we expect in the horizontal direction
$\sigma^\text{BS}_x \approx 65\mum$, $\sigma'_x \approx 2.1\,\mu\text{rad}$,
$\sigma_{p_x} \approx 14\MeV$, and in the vertical direction
$\sigma^\text{BS}_y \approx 95\mum$, $\sigma'_y \approx 1.5\,\mu\text{rad}$,
$\sigma_{p_y} \approx 10\MeV$. Both the measured horizontal and vertical
initial beam spot width values are close to their expectations (70 and 90\mum).

\begin{table}[bh]

\centering

 \topcaption{The beam parameters and related quantities at the CMS interaction
point for the $\beta^* = 90\unit{m}$ run at $\sqrt{s}=13\TeV$ in 2018.}

 \label{tab:beam}

 \begin{tabular}{lc}
 \hline
 Beam parameter & Value \\
 \hline
  Horizontal amplitude function at IP, $\beta^*_x$ & 45\unit{m} \\
    Vertical amplitude function at IP, $\beta^*_y$ & 90\unit{m} \\
  Normalised emittance (expected), $\varepsilon_n$ & $1.2\text{--}1.6\mum$ \\
  Horizontal beam spot width (measured), $\sigma^\text{BS}_x$ &
                                         $70\text{--}110\mum$ \\
    Vertical beam spot width (measured), $\sigma^\text{BS}_y$ &
                                         $90\text{--}110\mum$ \\
  Half of horizontal crossing angle & $60\,\mu$rad \\
 \hline
 \end{tabular}

\end{table}

\begin{figure}[b]
 \centering
 \includegraphics[width=0.9\textwidth]{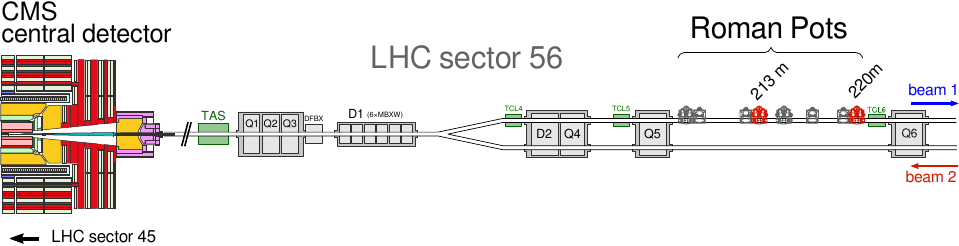}

 \caption{Plan view of LHC section 56 (Arm 2) close to CMS (point 5 at the
LHC). The location of the near and far Roman pots of Arm 2 are indicated in
red. Also, various magnetic beam elements (dipoles D1 and 2, quadrupoles Q1
to Q6) as well as other Roman pots not used for this paper are shown.}

 \label{fig:lhc_totem}

\end{figure}

\subsection{Roman pot detectors}
\label{sec:rp}

\begin{figure}[b]

 \centering
 \includegraphics[width=0.58\textwidth]{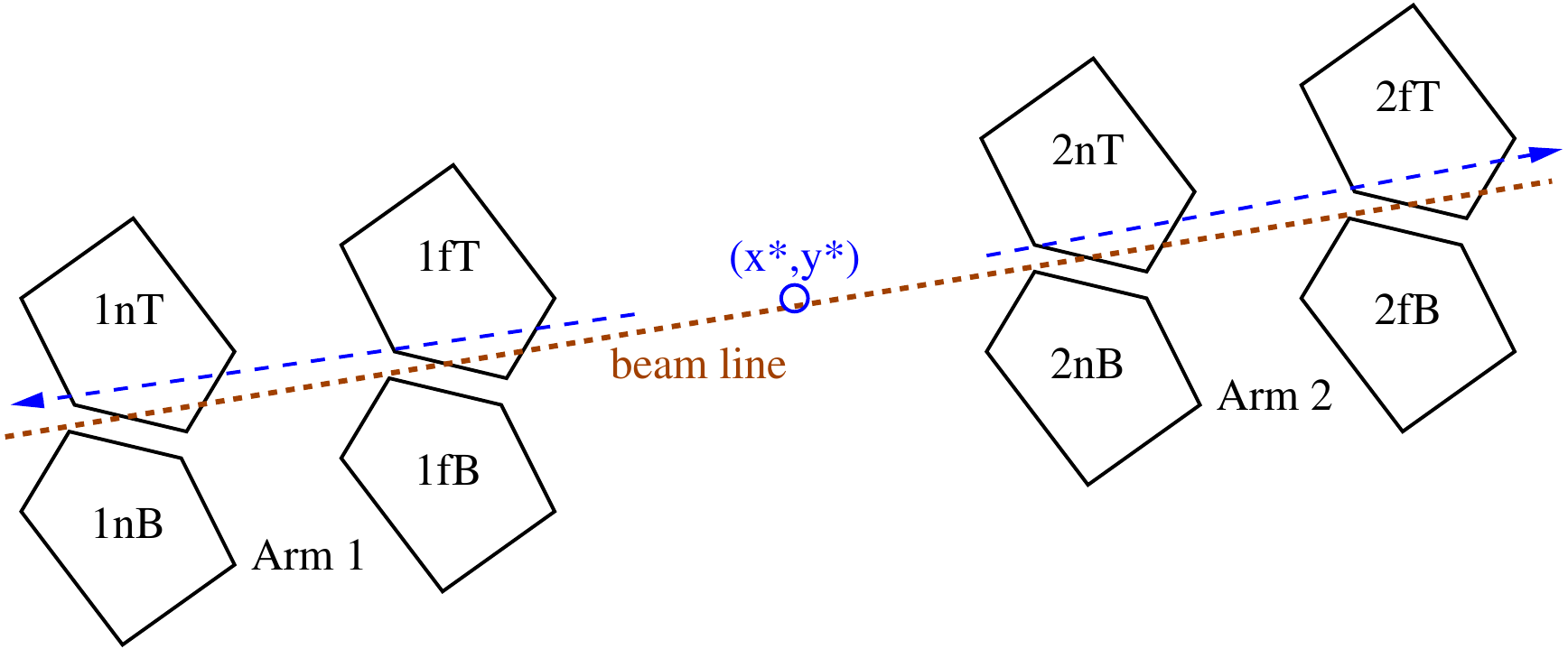}
 \hspace{0.02\textwidth}
 \includegraphics[width=0.38\textwidth]{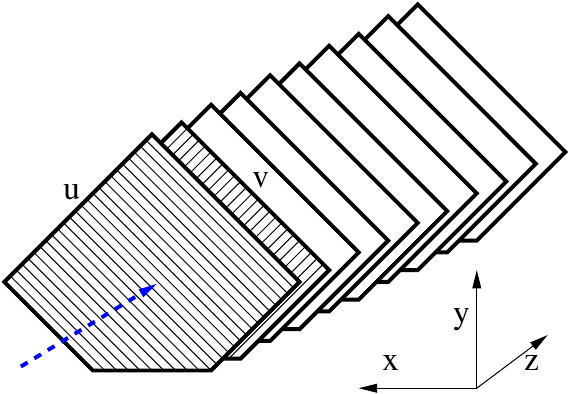}

 \caption{Left: The eight RP units in Arm 1 and Arm 2 with the beam line, along
with the location $(x^*,y^*)$ of the primary interaction and two proton
trajectories (dashed blue arrows), not to scale. Right: An RP unit comprising
10 layers with strips oriented alternately in the $u$ and $v$ directions. The
blue dashed arrow represents an incoming proton. The axes of the local
coordinate system are indicated with black solid arrows.}

 \label{fig:totem_rp}

\end{figure}

The intersection points of the LHC are numbered from 1 to 8. The CMS detector
operates at interaction point 5, and the adjacent LHC sectors are referred to
as ``45'' and ``56''. The proton spectrometer of the TOTEM experiment consists
of two telescope arms, referred to as ``Arm 1'' (in sector 45) and ``Arm 2''
(in sector 56) for positive and negative pseudorapidities,
respectively~\cite{Anelli:2008zza,Antchev:2013hya}. In each arm there are two
stations located at about $\pm 213\unit{m}$ (near) and $\pm 220\unit{m}$
(far) relative to the nominal IP (Fig.~\ref{fig:lhc_totem}). Each station
consists of two units. Only the units with vertical pots are considered in
this paper. Each unit has two RPs, one located above (``top'' or T), and one
below (``bottom'' or B) the LHC beam.  The names of the RP detector layer
groups are given in Table~\ref{tab:notations} and Fig.~\ref{fig:totem_rp}
(left panel). The term ``parallel'' refers to cases where both RP detectors
above or below the beamline (top-top or TT, bottom-bottom or BB) have signals
above predefined thresholds, whereas the ``diagonal'' configuration refers to
the other two cases (top-bottom or TB, bottom-top or BT).

An RP contains ten layers of silicon strip detectors, placed alternately in two
orthogonal orientations (u and v) with five layers in each. The strips are
directed at $45^\circ$ with respect to the vertical ($y$) axis for the far pots
(Fig.~\ref{fig:totem_rp}, right), and at $37^\circ$ or $53^\circ$ for the near
pots. The layers can approach the beam at a distance of a few millimetres
without affecting the LHC operation (they are retracted before each beam
injection and inserted again afterwards).

\begin{table}

\centering
 \topcaption{The naming of the various RP layer groups.}

 \label{tab:notations}

 \begin{tabular}{ccccl}
  \hline
  \multirow{2}{*}{Arm} &
  \multirow{2}{*}{Location} & Vertical &
  \multirow{2}{*}{Proj.} & \multirow{2}{*}{Name} \\
    & & position & & \\
  \hline
  1 & near & top    & $u$ & 1nTu \\
  1 & near & top    & $v$ & 1nTv \\
  1 & near & bottom & $u$ & 1nBu \\
  1 & near & bottom & $v$ & 1nBv \\
  1 & far  & top    & $u$ & 1fTu \\
  1 & far  & top    & $v$ & 1fTv \\
  1 & far  & bottom & $u$ & 1fBu \\
  1 & far  & bottom & $v$ & 1fBv \\
  \hline
 \end{tabular}
\quad
 \begin{tabular}{ccccl}
  \hline
  \multirow{2}{*}{Arm} &
  \multirow{2}{*}{Location} & Vertical &
  \multirow{2}{*}{Proj.} & \multirow{2}{*}{Name} \\
    & & position & & \\
  \hline
  2 & near & top    & $u$ & 2nTu \\
  2 & near & top    & $v$ & 2nTv \\
  2 & near & bottom & $u$ & 2nBu \\
  2 & near & bottom & $v$ & 2nBv \\
  2 & far  & top    & $u$ & 2fTu \\
  2 & far  & top    & $v$ & 2fTv \\
  2 & far  & bottom & $u$ & 2fBu \\
  2 & far  & bottom & $v$ & 2fBv \\
  \hline
 \end{tabular}
\end{table}

\begin{figure}[t]
 \centering
 \includegraphics[width=0.45\textwidth]{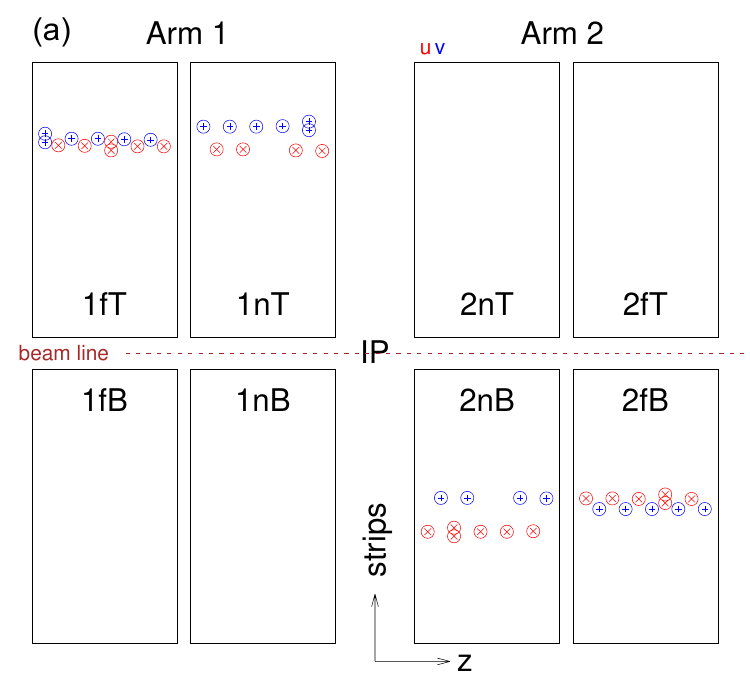}
 \includegraphics[width=0.45\textwidth]{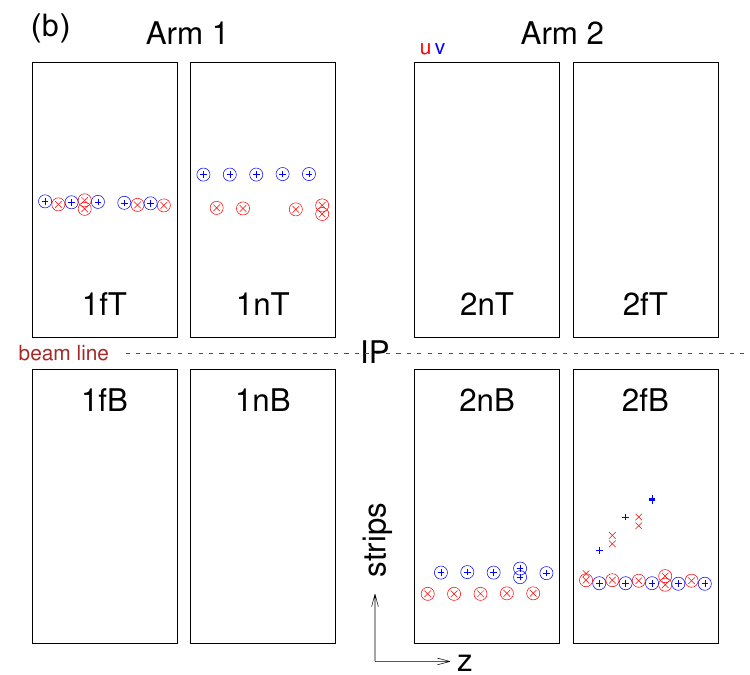}
 \includegraphics[width=0.45\textwidth]{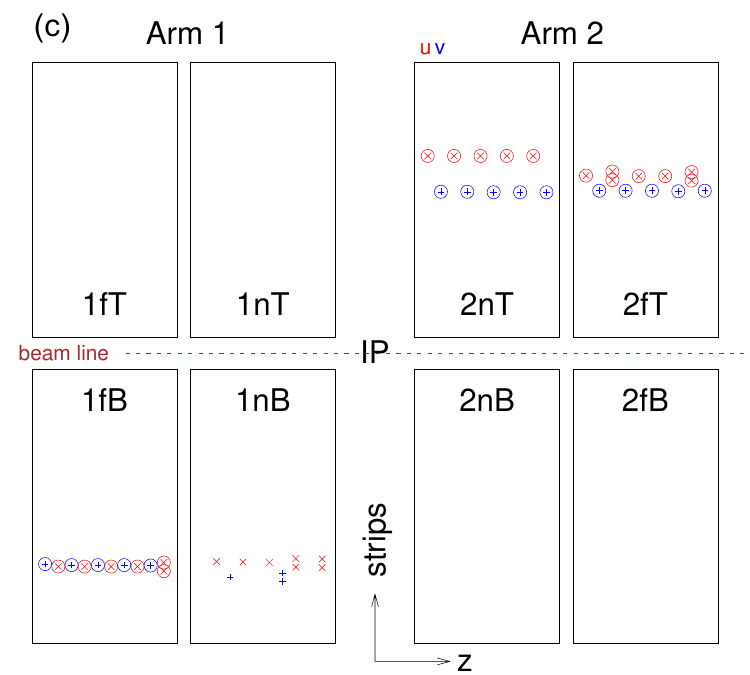}
 \includegraphics[width=0.45\textwidth]{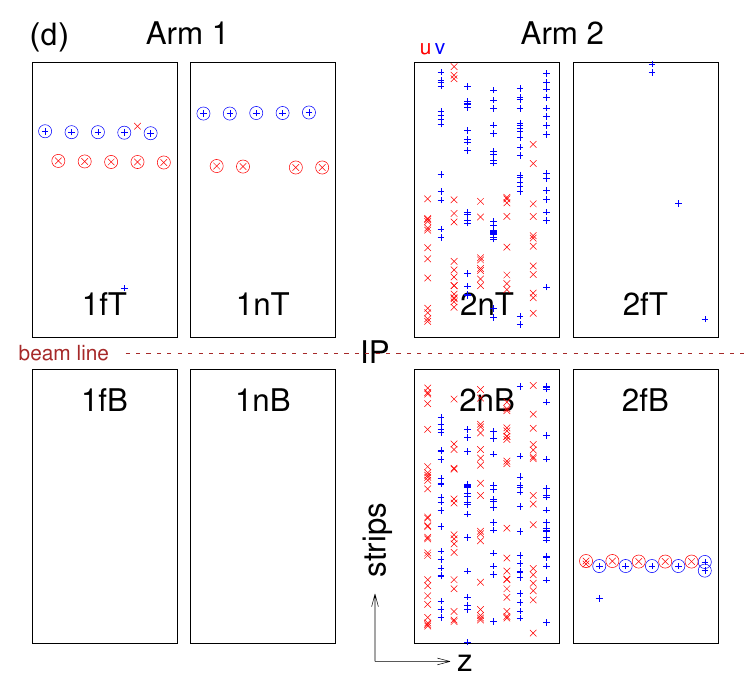}
 \caption{Schematic displays of four events with hits and projections of
reconstructed local tracks (tracklets) in RPs, with outlines of Arm 1 and 2,
near and far, upper and lower plots (not to scale). Strips of reconstructed
clusters in $u$ and $v$ layers are plotted with red {x} and
blue {+} markers, respectively. Two vertically adjacent symbols
represent a two-strip cluster. Reconstructed clusters on a found tracklet are
marked with circles. From left to right and from top to bottom: (a) normal
event, (b) normal event with additional secondary particles; (c) not
reconstructed event (less than three hits in the $v$ orientation of 1n); (d) not
reconstructed event (hadronic interaction in some beam element before 2n).}

 \label{fig:events}

\end{figure}

Schematic displays of some events with hits and projections of reconstructed
local straight tracks, called ``tracklets'', in RPs are displayed in
Fig.~\ref{fig:events}. Normally, the scattered protons are detected by both the
near and far pots, yielding a total of four tracklets. Events with more than
four tracklets, pointing to energetic secondaries or multiple simultaneous
primary pp collisions, are excluded from the analysis. The RPs share a common
readout with the CMS detector.

\clearpage

\subsection{Proton trajectory}
\label{sec:initial}

Once the scattered protons leave the interaction region, they traverse the
magnetic fields created by the CMS solenoid and the LHC beam magnets until
most of them reach the first and then the second RP stations. During their
flight, such protons do not traverse the beam pipe. The RP stations have
their own secondary vacuum, separated by a $300\mum$ thin steel window from
the primary LHC vacuum, and they are in a region where the magnetic field is
zero. Of course, to determine the scattered proton kinematics the
inhomogeneous magnetic fields due to the sequence of several quadrupole and
dipole magnets need to be modelled.

The effects of multiple scattering and nuclear collisions in the detector
material and in the stainless steel window that separates the LHC primary
vacuum from the RP secondary vacuum, need to be considered. The sensitive
detector elements are silicon layers with a thickness $d = 300\mum$. Silicon
(steel) has a radiation length $X_0 = 9.370$ ($1.757$) \cm, and a nuclear
collision length $\lambda = 30.16$ ($10.37$) \cm. For multiple scattering
within one layer, the standard deviation of the scattering angle, in a
Gaussian approximation~\cite{ParticleDataGroup:2024cfk}, is given by $\theta_0
\approx 13.6\MeV/(\beta c p) \times z \sqrt{d/X_0}$, where $z$ is the charge number of
the incident particle, and we have neglected the logarithmic correction. With
the proton momentum $p = 6500\GeV$, we get $\theta_0 \approx 0.12$ ($0.27$)
$\mu\text{rad}$ for the silicon (steel) layer. Within an RP, assuming 9
detector planes, after a distance of 36\mm between the first and the last
layer, the average resulting shift is 15\unit{nm}, and is neglected. After 10
scattering planes and a distance of 7\unit{m} between two pots, the shift is
3\mum, which again is neglected. The probability of a nuclear collision within
a silicon (steel) layer is $d/\lambda \approx 10^{-3}$ ($3\times 10^{-3}$). In
the case of an RP with two thin windows and 10 layers, it adds up to 1.6\%. The
nuclear collisions thus contribute to the tails of the hit location
distributions.

\subsection{The data set}

In the high-$\beta^*$ data-taking period considered here, the average number of
pp collisions per bunch crossing is small (0.1--0.3), and for the
same reason the collected integrated luminosity is limited. For this data set,
the Level 1 (hardware) trigger requires detected protons in the far RPs in each
arm, in parallel or diagonal configurations (Section~\ref{sec:rp}). Triggering
is based on ``trigger roads'', groups of 32 consecutive silicon strips. The
trigger bit is set if trigger roads at the same location have signals above a
predefined threshold in several detector layers. The elastic pp cross section
is much larger than that of central exclusive production events, and therefore
the available bandwidth of the data acquisition system would be saturated.
Suitable combinations of the trigger bits from the RPs are used to reject
elastic events. The high-level trigger has multiple
components~\cite{CMS:2016ngn}; the pixel and track activity filters of the CMS
silicon tracker system require hits or tracks from centrally produced charged
particles.

For these studies two data sets are used. One data set with no centrally
reconstructed charged particles (``0-track'' data set with some unrelated hits
in the pixel detector) contains 39 million events of which 29 million have
all four tracklets reconstructed. The other data set has two oppositely-charged
centrally reconstructed particles (``2-track'' data set) contains 119 million
events of which 88 million have all four tracklets reconstructed.

\section{Tracklet reconstruction}
\label{sec:tracklet}

In various exclusive physics analyses, the classification of signal events is
based on the conservation of momentum between the incoming and outgoing
particles. To maximise the selection efficiency, an optimisation of
the standard local tracklet
reconstruction~\cite{Anelli:2008zza,Antchev:2013hya} was performed, which led
to improvements of resolution and uncertainty estimations.

\begin{figure}[b]

 \centering
 \includegraphics[width=0.66\textwidth]{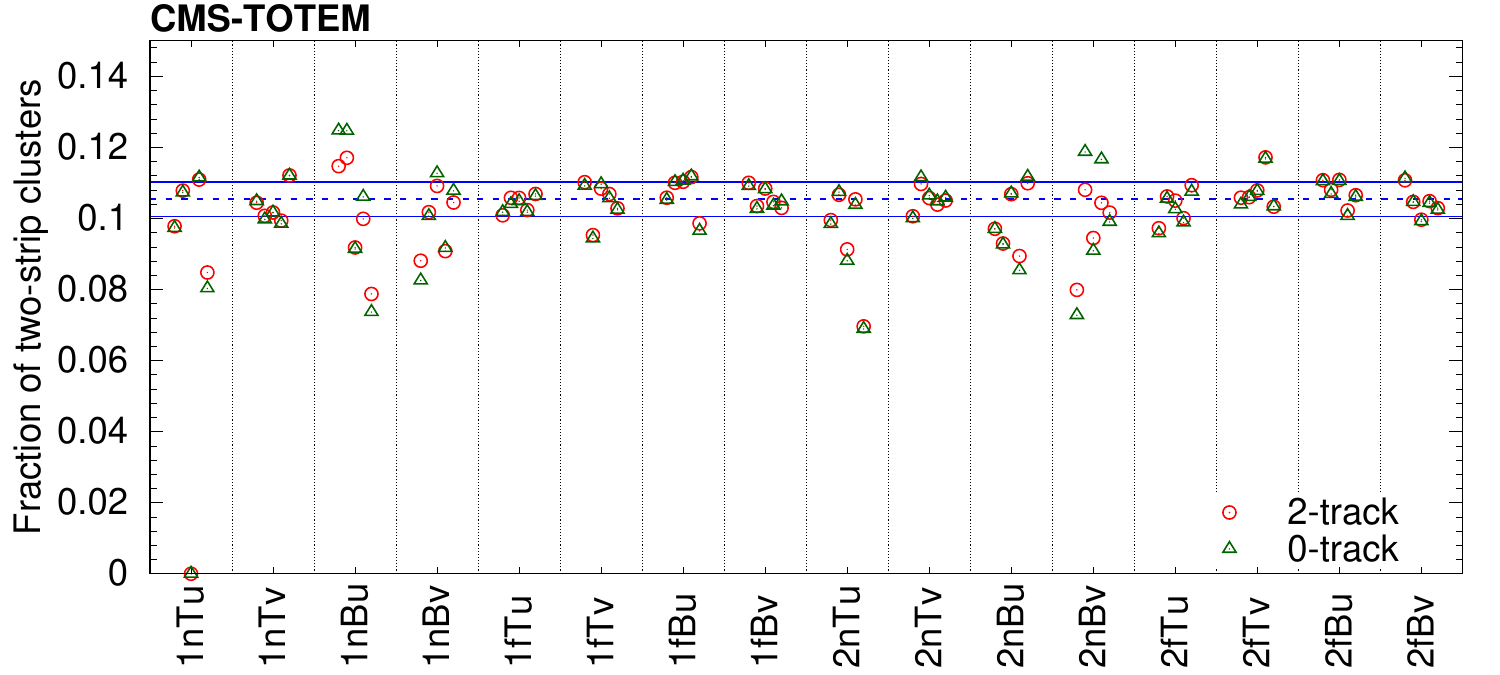}
 \includegraphics[width=0.33\textwidth]{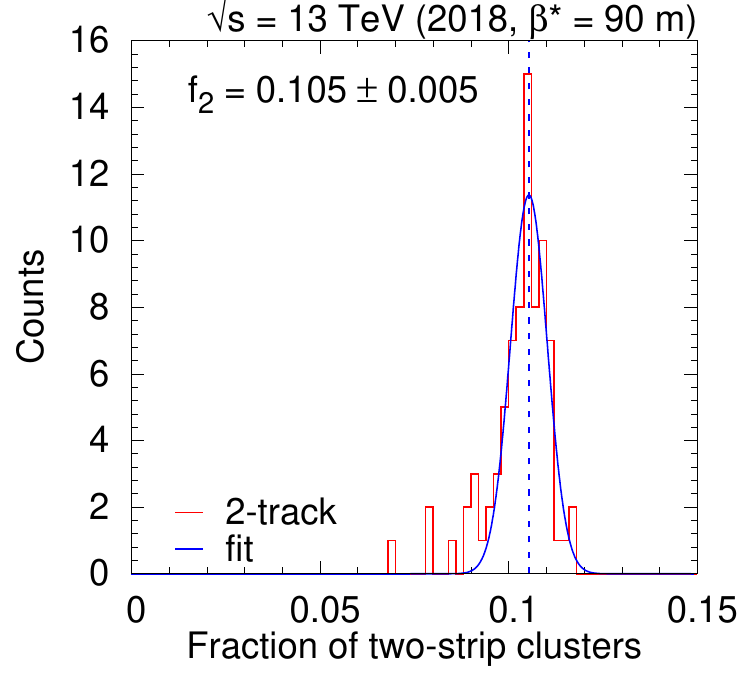}

 \caption{Left: Fraction of two-strip clusters for each detector layer, in
groups of similarly oriented layers. Right: Distribution of the fraction of
two-strip clusters with the result of a Gaussian fit.}

 \label{fig:2strip}

\end{figure}

Every RP detector layer contains 512 strips, each with a width (pitch) of
$66\mum$. A hit cluster is a group of adjacent strips, each with a collected
charge above a certain threshold. Most of the clusters are one-strip wide,
with the hit in the central area of the strip. The less frequent two-strip
clusters arise through drifting and diffusing electrons from hits in the
region between two strips (e.g. in panel (a) of Fig.~\ref{fig:events}, Arm 1,
1nT). The fraction of two-strip clusters for each detector layer is shown in
Fig.~\ref{fig:2strip}, in groups of similarly oriented layers (layer groups).
The result of a Gaussian fit is also indicated, with a mean $f_2 = 10.5\%$.
The outliers, mostly at smaller values, belong to layers with low strip-level
efficiencies. Two-strip clusters carry important information and provide
better spatial resolution than one-strip clusters
(Section~\ref{sec:polygon}). Clusters with three or more strips are rare,
with a rate below half a percent, demonstrating that the effect of capacitive
coupling between neighbouring strips is negligible.

Lacking an analog strip readout, the location of a cluster cannot be computed
using a charge-weighted cluster mean. Hence the location of a one-strip cluster
can be associated to an integer number $s_i$, whereas the location of two-strip
cluster is the average of their integer numbers (half-integer).

\subsection{Pattern recognition}
\label{sec:tracklet_fits}

In conventional applications, track finding usually starts with local hit
reconstruction that, in case of segmented detectors, also includes cluster
finding. Based on the hits found, track seeds (compatible hit multiplets) are
built. These are used as starting points for a global track finding by
extending them towards other detector layers and picking up free, unused hits
(through a Kalman filtering method~\cite{Fruhwirth:1987fm,Billoir:1990we}).
Finally, the track candidates are scanned for split or multiply reconstructed
trajectories, their momentum vectors are determined, and they are used to fit
a common global event vertex. The calibration of a tracking detector
includes its alignment by means of cosmic ray muons or particles originating
from particle collisions, but sometimes also implies a gain calibration. The
performance of a tracking system is usually characterised by its position and
momentum resolutions~\cite{Fruhwirth:2020zbo}.

In the case of RP layers, local hit reconstruction is performed by searching
for groups of contiguous strips, mostly one-strip and two-strip clusters,
resulting in integer and half-integer cluster coordinates. The strips do not
provide pulse height information, but only binary information based on signal
discrimination. Since a cluster is a series of adjacent strips with a given
width, the trajectory-hit residuals follow rectangular distributions. For this
reason, instead of performing least-square fits, a different approach is chosen
with the aim to use as much information as possible.

In each RP, the orientations $u$ and $v$ are handled separately, since the
strips in the two orientations are orthogonal. The tracklet model is a straight
line with local hit coordinate $u_i$ (or $v_i$) in the $i$th layer:
\begin{equation}
 u_i = a z_i + b + \delta_i,
\end{equation}
where $a$ is the slope and $z_i$ is the coordinate of the location of the
detector layer along the beam direction. The centre of the RP is at $z = 0$
where the tracklet has the coordinate $b$ (intercept). The misalignment of the
detector layers is quantified by relative shifts $\delta_i$
(Section~\ref{sec:layer_shifts}). Common translation (all $\delta_i$ values are
equal) and shear (linear relation $\delta_i = \tau_\text{shear} z_i$) are weak
modes and are difficult to detect, and therefore the alignment is carried out
with the outermost (first and fifth) layers fixed, $\delta_1 = 0$ and $\delta_5
= 0$. The $z$-locations of the layers for Arm 1 are $20.3$, $11.3$, $2.3$,
$-6.7$, $-15.7\mm$ ($u$ orientation) and $15.7$, $6.7$, $-2.3$, $-11.3$,
$-20.3\mm$ ($v$ orientation); the same values hold for Arm 2, but with opposite
sign.

An ideal track traverses two pots (near and far), \ie, in total four layer
groups, and each layer group can have up to five hits. Using all events in the
0-track and 2-track data sets, a database of hit location patterns is built.
This database is essential for pattern matching, in connection with a
tag-and-probe method~\cite{CMS:2010svw}, for precise relative alignment
(Section~\ref{sec:layer_shifts}) and for the determination of individual strip
hit efficiencies (Section~\ref{sec:stripEffic}). Such a procedure employs an
unbiased sample of probe objects that is used to measure the efficiency of a
specific set of selection criteria. Here tracklets with at least four hits are
tagged, and in each layer a hit at the expected location is searched for. In
the case of layer group 1nTu, at least three hits are required since one of its
layers is inefficient.

To greatly reduce the number of hit location patterns, common translation (of
the first layer, $s_1$) and average shear are removed. Since we want to
distinguish between one-strip and two-strip clusters, only integer
displacements $- [s_1]$ are applied. This way, the relative location in the
first layer will be either 0 (one-strip) or 0.5 (two-strip). If the first layer
is empty, the next layer with valid hit is taken as reference. The average
integer shear is also removed by subtracting the value $i \langle s_i -
[s_1]\rangle$ for layer $i$, where $\langle s_i - [s_1] \rangle$ is deduced
from $\sum i (s_i - [s_1]) / \sum i$ by rounding down to an integer, where
$s_i$ is the strip number on layer $i$. Some examples for the final patterns
are $(0, 0.5, 1, 1, 2)$, $(0, 0, -1, -3, -2)$, or $(0.5, 1, 1, 2.5, 2)$. Our
data sets provide a database of relative hit location patterns with about 15
thousand entries for each layer group (there are altogether 16 layer groups).

\begin{figure}[ht]

 \centering
 \includegraphics[width=0.32\textwidth]{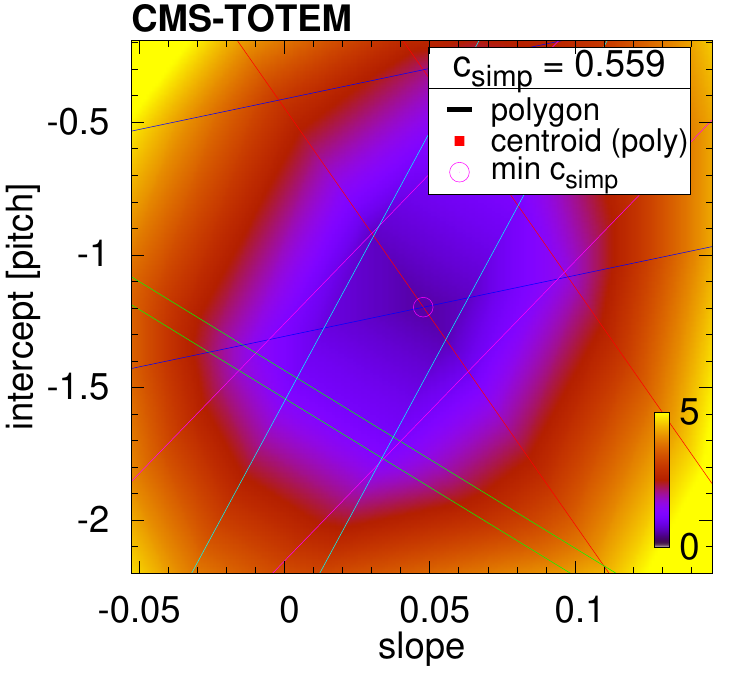}
 \includegraphics[width=0.32\textwidth]{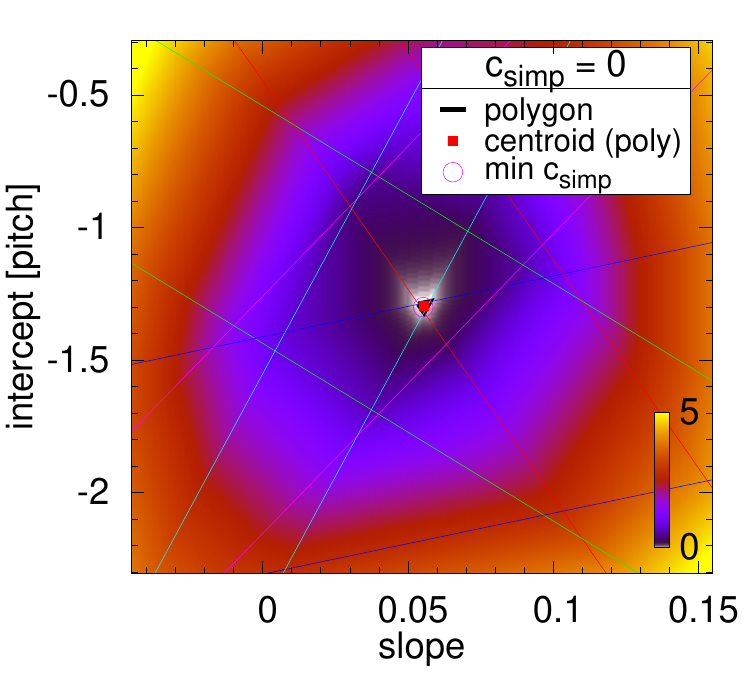}
 \includegraphics[width=0.32\textwidth]{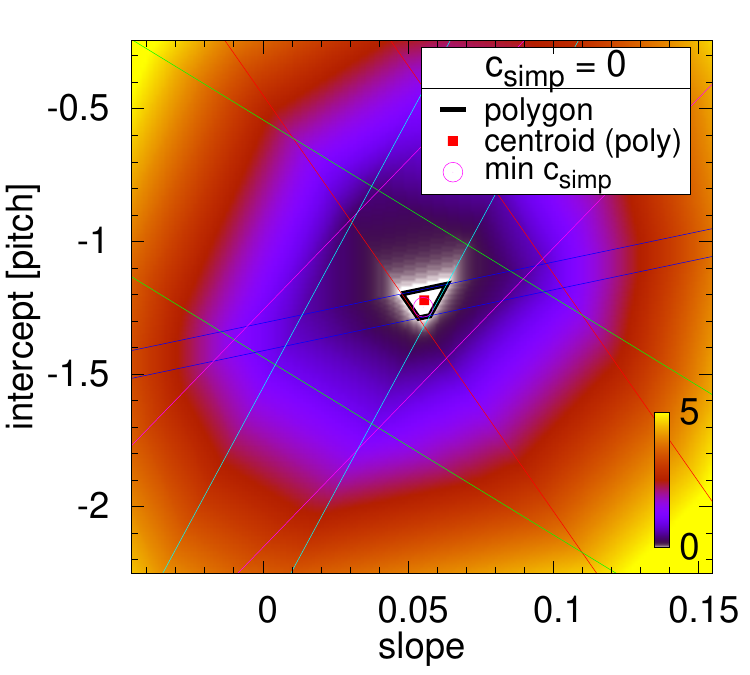}

 \includegraphics[width=0.32\textwidth]{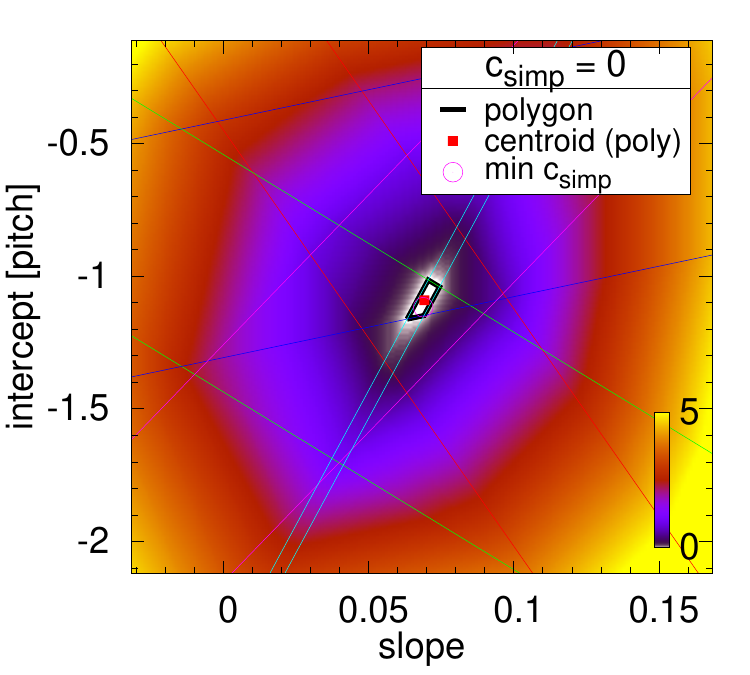}
 \includegraphics[width=0.32\textwidth]{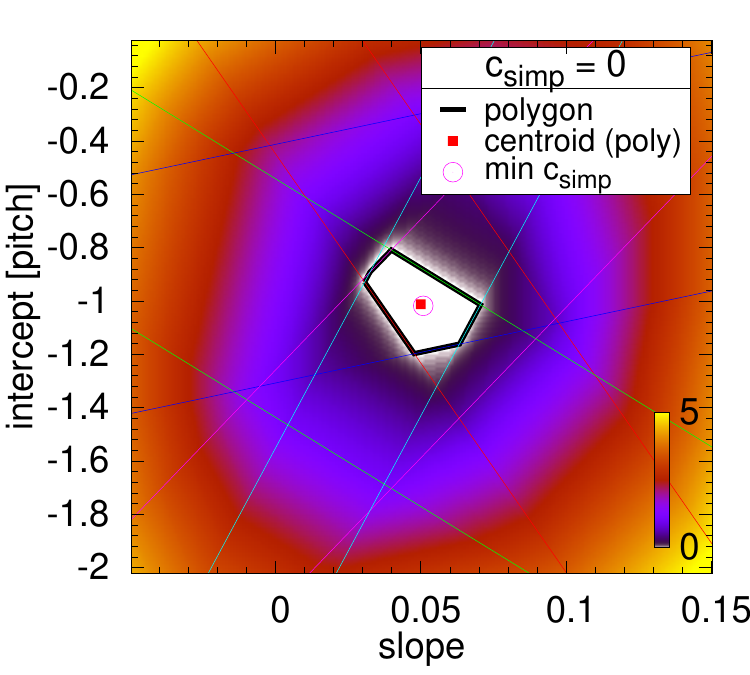}
 \includegraphics[width=0.32\textwidth]{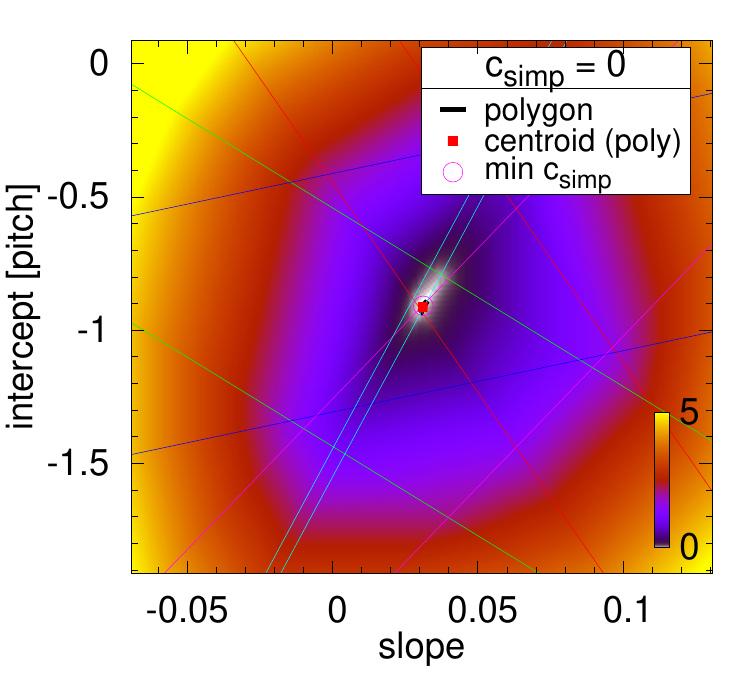}

 \caption{Examples of tracklet fits. The colour represents the value of the
tracklet penalty function $c$ in the intercept-slope ($b\text{--}a$) plane.
Bands corresponding to individual detector layers are shown with differently
coloured parallel straight lines. The intersection of these bands, if it
exists, is shown as a white polygonal area ($c_\text{simp} = 0$) framed with
thick black lines, and its centroid is marked with a red filled square. If
such intersection does not exist (top left subfigure), the result of the
simplex minimisation $c_\text{simp} > 0$ is given, and the location of the
minimum indicated with a purple open circle.}

 \label{fig:tracklet_fits}

\end{figure}

\label{sec:polygon}

If the relative shifts $\delta_i$ are known, the predicted hit location on the
$i$th layer is $u^\text{pred}_i = a z_i + b + \delta_i$ in strip-width units.
The hit distribution of the incoming protons is locally uniform. The incidence
angles of the protons are very small, and the protons enter the silicon planes
perpendicularly to the planes. If the proton hit is closer than $w$ to the
boundary line between two adjacent strips, it deposits enough energy, such that
both strips collect charge above threshold: we get a two-strip cluster. Here
$w$ is measured in pitch units. If the hit is farther than that, the proton
deposits a sizeable amount of charge only in one central strip, leading to a
one-strip cluster. Because of the local uniformity of the incoming proton
distribution, the occurrence of two- and one-strip clusters is directly
proportional to the width of the areas described above. For two-strip clusters
$f_2 = 2 w$, hence $w \approx 0.0525$. For one-strip clusters the cluster
centre $u^\text{meas}$ is expected to lie closer than $1/2 - w = 0.4475$ to the
hit, whereas for two-strip clusters the distance should be below $w$. This
way, two-strip clusters provide better spatial resolution by a factor of $(1 -
f_2)/f_2 \approx 8.5$ with respect to one-strip clusters. The above
requirements can be written down as
\begin{gather}
 \abs{u^\text{meas} - u^\text{pred}} < w
\qquad
\text{or}
\qquad
 u^\text{meas}_i - a z_i - \delta_i - w < b <
 u^\text{meas}_i - a z_i - \delta_i + w.
\end{gather}
These two inequalities define a band in the intercept-slope ($b\text{--}a$)
plane (Fig.~\ref{fig:tracklet_fits}).

Since each detector layer provides such a constraint, we look for the
intersection of five bands, \ie, a polygonal area. The determination of this
common area is carried out iteratively. First, the intersection of the first
two bands, \ie, a parallelogram, is determined. Next, the third band is taken,
and intersections of its boundary lines with the sides of the parallelogram are
looked for. If such points are found, they are added to the polygon as new
vertices, and the vertices that became unnecessary are removed. The
process is continued until all bands are dealt with, and we end up with a
polygon $(a_j,b_j)$ of $j = 1, \dots, n$ vertices. All points within this
final polygon are equally probable and valid. The best value and variance of
the intercept are represented through the centroid and the moment of inertia of
the polygon~\cite{FGBV-96-05}. Some examples for the common polygons are shown
in Fig.~\ref{fig:tracklet_fits}.

\begin{figure}

 \centering
 \includegraphics[width=0.9\textwidth]{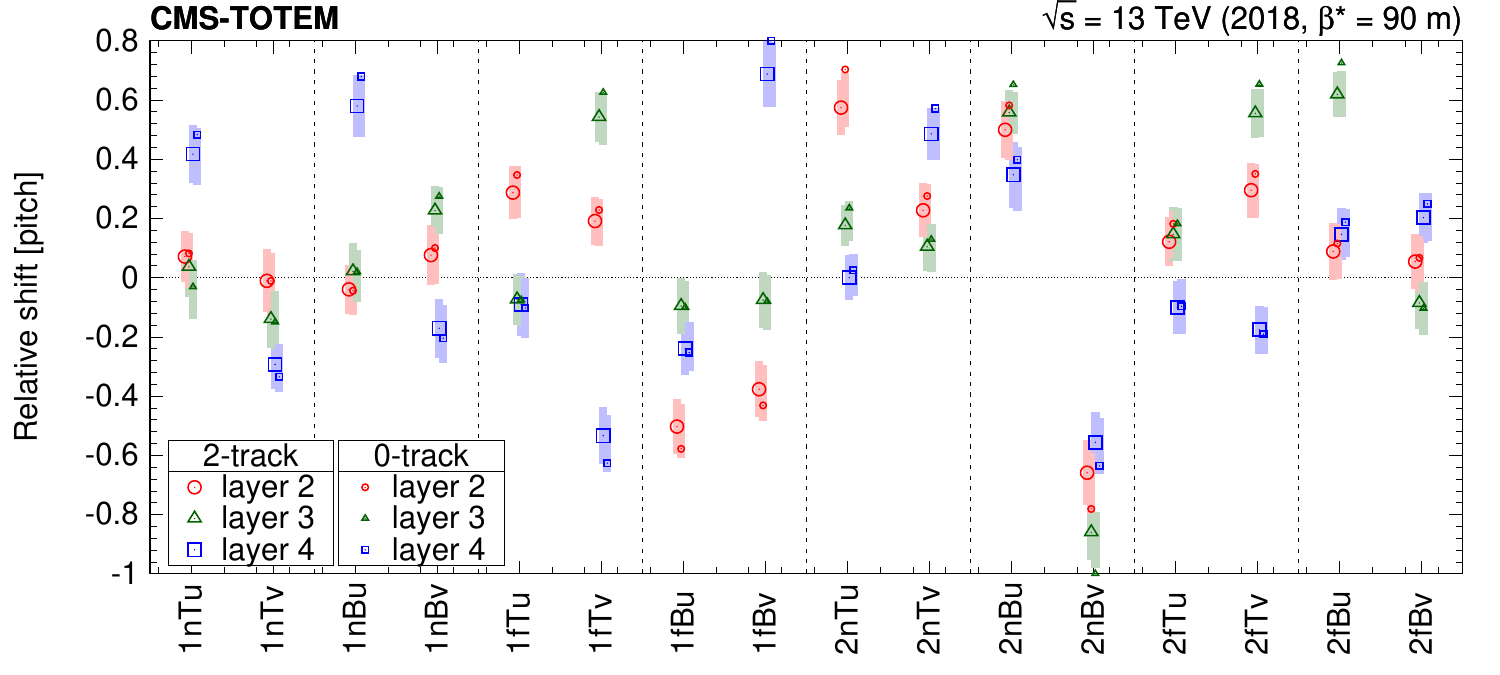}

 \caption{Deduced alignment parameters for inner layers ($\delta_2$ -- red
circle, $\delta_3$ -- green triangle, $\delta_4$ -- blue square) in layer
groups, in units of strip-width. Larger open symbols represent the 2-track data
set, and the smaller ones are based on the 0-track data set. Shaded bars
indicate the estimated systematic uncertainties
(Section~\ref{sec:layer_shifts}).}

\label{fig:planeShifts}

\end{figure}

In some cases such a common polygon does not exist. This can happen if there
are one or more layers where the cluster centre is farther than the expected
distance $w$ from the location of the predicted hit. In such cases the above
polygon method does not work: one needs to define a suitable goodness-of-fit
measure. It is accomplished by means of a penalty function $c$: if the
distance $d$ of the predicted hit and the cluster is larger than $w$, we
collect and sum terms $\abs{d-w}$. Such a choice, instead of the squared
difference, favours the suppression of outlier (noise or background) clusters.
The minimisation of the $c$ value in the $b\text{--}a$ plane is accomplished by
using the downhill simplex method of Nelder--Mead~\cite{Nelder:1965zz}. The
minimisation usually converges in 3-5 steps, but always finds the minimum in
less than 50 steps. Some examples can be seen in Fig.~\ref{fig:tracklet_fits}.
In these cases the standard deviation of the measurement in the $b$ direction
is taken to be 0.3 units. This choice ensures a smooth and monotonic relation
between the measured and the predicted variances of the proton momentum sum
($\sum p_x$) distributions (Section~\ref{sec:results},
Fig.~\ref{fig:resolution_x}).

\begin{figure}[th]

 \centering
 \includegraphics[width=\textwidth]{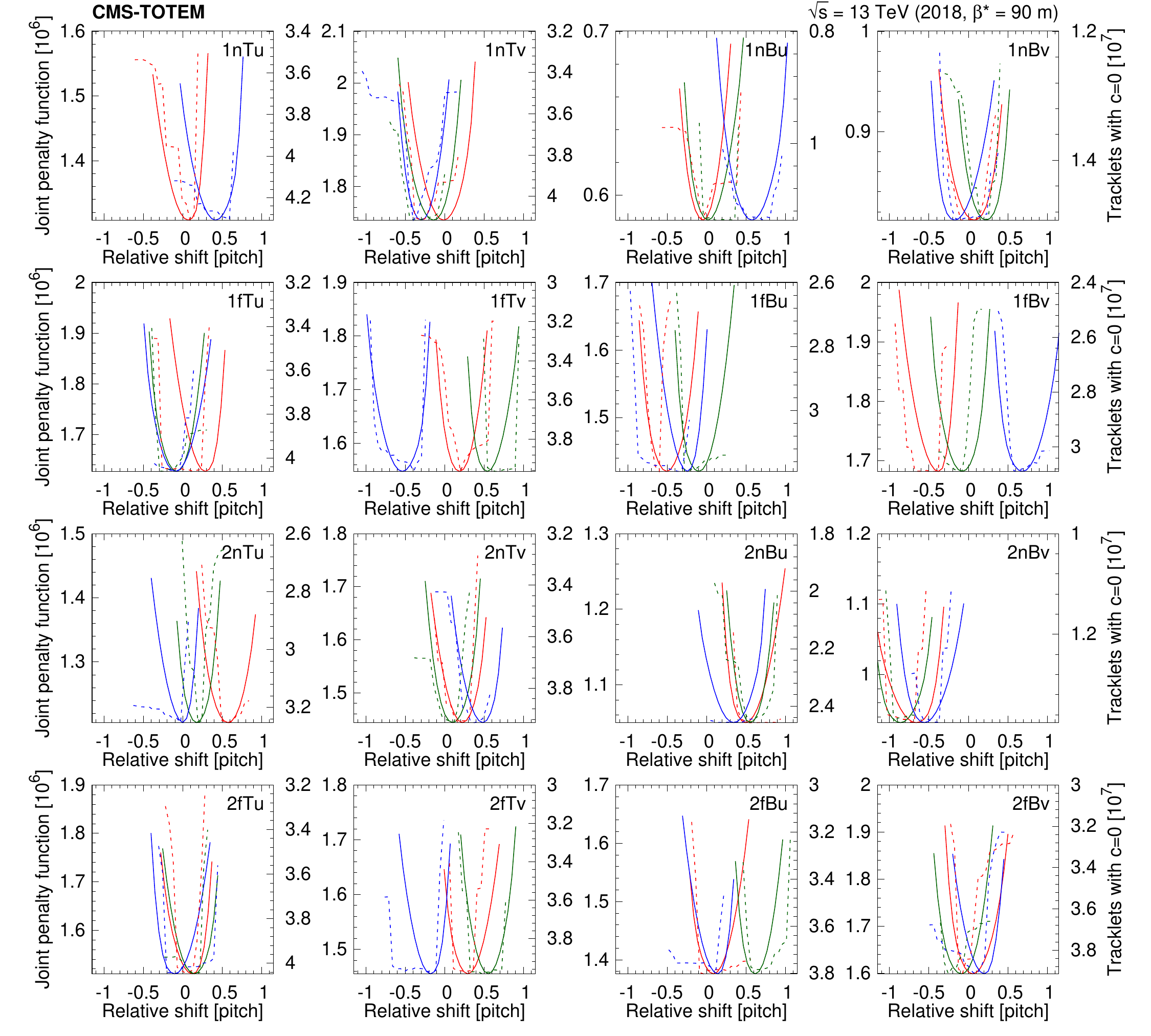}

 \caption{One-dimensional line scans of the relative shifts ($\delta_2$ -- red,
$\delta_3$ -- green, $\delta_4$ -- blue) for the 16 layer groups around the
best values found. The joint $c$ value is plotted as a function of the relative
shifts of the inner layers around the found minima (solid curves, left vertical
scale). Goodness-of-fit from an alternative method, counting tracks with $c =
0$, is also indicated (dashed lines, right vertical scale, number of such
tracks with inverted axis). In the 1nTu panel (upper left corner) the curve
corresponding to the third layer is missing since that layer is inefficient.}

 \label{fig:planeShifts_line}

\end{figure}

\subsection{Relative alignment of the detector layers}
\label{sec:layer_shifts}

Tracklet fits can be converted into a tool to determine the relative alignment
of the detector layers. The parameters of this three-dimensional problem are
the relative shifts of the inner layers $(\delta_2, \delta_3, \delta_4)$. The
goal is to minimise the joint goodness-of-fit measure, \ie, the sum of $c$
values for a large set of the tracklets. The employed minimisation method is
again the downhill simplex method of Nelder--Mead~\cite{Nelder:1965zz}. The
best $\delta_i$ values for the 16 layer groups, corresponding to the lowest
sum, are displayed in Fig.~\ref{fig:planeShifts}. The indicated systematic
uncertainties represent the region where the sum of the $c$ values (tracklet
penalty) would increase by one percent, a well-visible change. The values from
the 2-track and 0-track data sets are compatible with each other. The standard
deviation of the relative shifts is about 0.35, in units of strip-width. The
deduced relative shifts are cross-checked with an alternative method by
counting the number of tracklets with $c = 0$, giving compatible results.
One-dimensional line scans around the minima are shown in
Fig.~\ref{fig:planeShifts_line}.

\begin{figure}[th]

 \centering
 \includegraphics[width=0.32\textwidth]{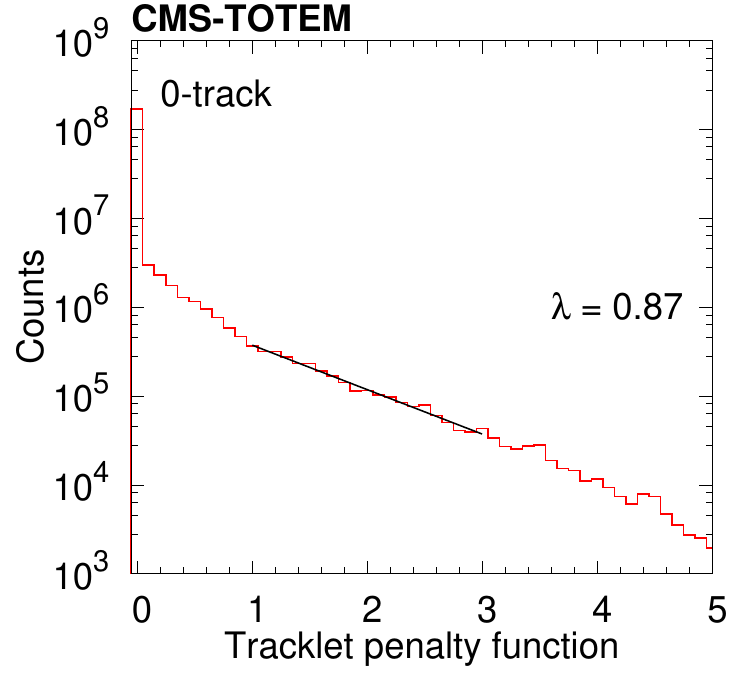}
 \includegraphics[width=0.32\textwidth]{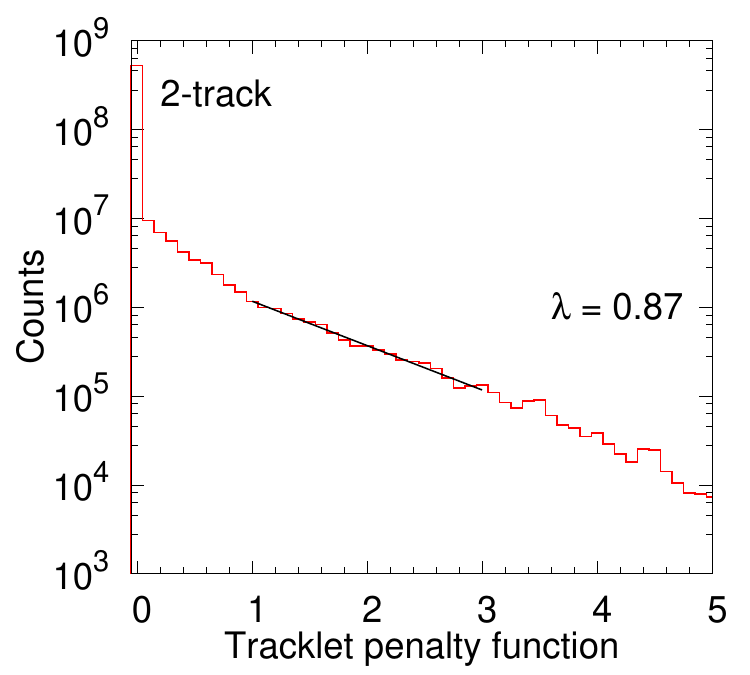}
 \includegraphics[width=0.32\textwidth]{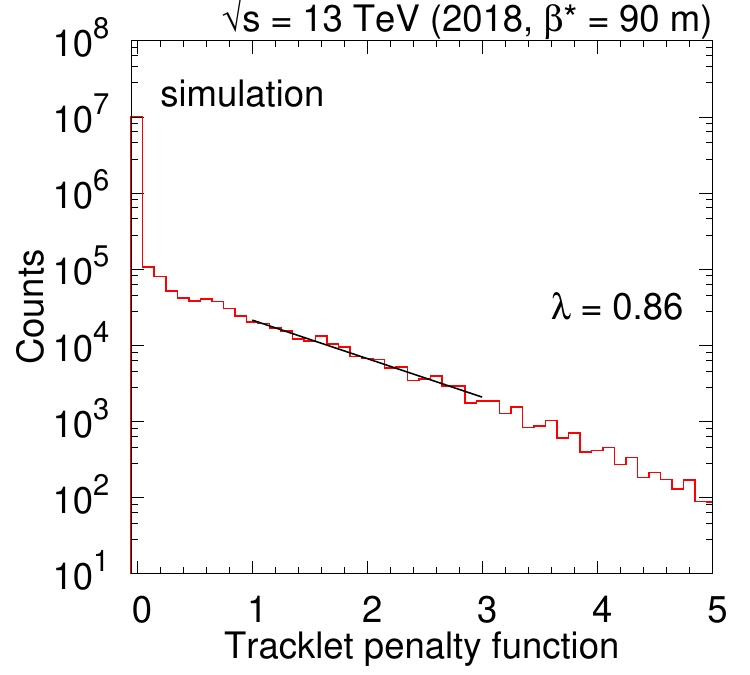}

 \vspace{0.01\textwidth}
 \includegraphics[width=0.32\textwidth]{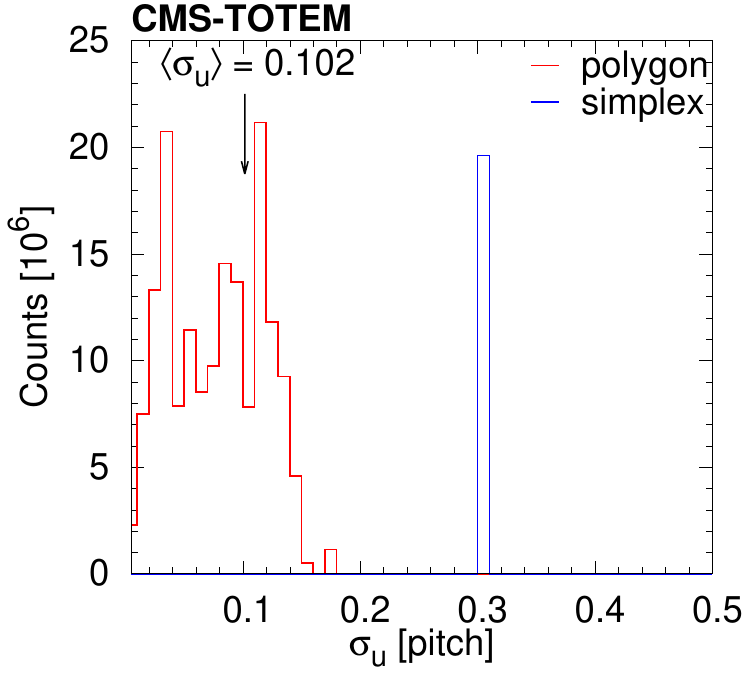}
 \includegraphics[width=0.32\textwidth]{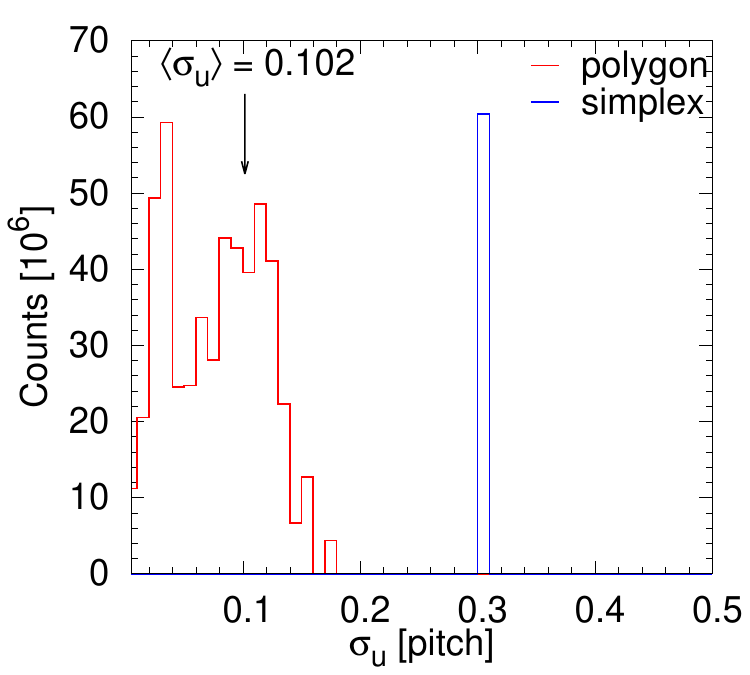}
 \includegraphics[width=0.32\textwidth]{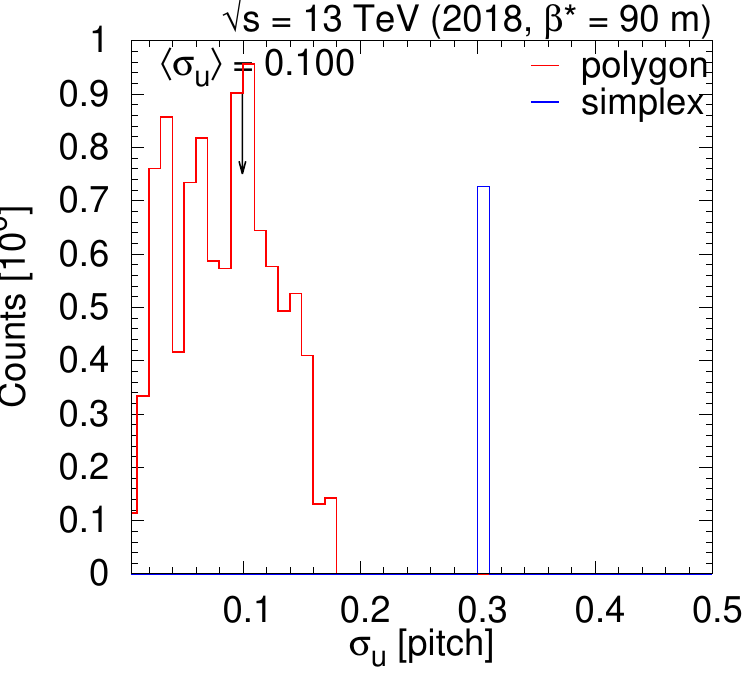}

 \caption{Upper: Distribution of the tracklet penalty function with the slope
$\lambda$ of an exponential fit (black line) in the range 1--3.
Lower: Distribution of the standard deviation of the fitted hit location in
strip-width units for the polygon method (red), and for those cases where the
simplex minimisation was necessary (blue, at 0.3). The three columns correspond
to the 0-track (left), 2-track (centre), and the simulated (right) data sets.
The vertical arrow indicates the location of the average value.}

 \label{fig:patterns_res}

\end{figure}

The distribution of the joint $c$ values for tracklets is plotted in the upper
panel of Fig.~\ref{fig:patterns_res}. Most tracklets (89\%) have $c = 0$, The
distribution of the standard deviation $\sigma_u$ of fitted hit location in
strip-width units for the polygon method, and for the cases where the simplex
minimisation was necessary, is shown in the lower panel of
Fig.~\ref{fig:patterns_res}. The average spatial resolution is 0.10 units, \ie,
about $6\text{--}7\mum$. This is to be compared with the naive expectation
from averaging three box distributions, $66\mum / \sqrt{12 \times 3} \approx
11\mum$, which worsens once the contribution from the standard deviation of the
relative shifts ($66\mum \times 0.35/\sqrt{3} \approx 13\mum$) is also
included.

To control and cross-check the deduced calibrations and to extract the
tracklet efficiencies, a fast but realistic simulation containing one million
events was set up. Tracklets are generated following a straight-line model
including the non-Gaussian nature of multiple scattering, as well as the
effects of nuclear collisions in the detector material. Hit creation
follows the measured fraction of two-strip clusters, and takes into
account the individual shifts and efficiencies. Conditions for triggering
and for tracklet reconstruction are properly included. A reasonable agreement
between the simulated and measured tracklet joint penalty and $\sigma_u$
distributions (Fig.~\ref{fig:patterns_res}, right column) is achieved.

In summary, the advanced treatment just described results in better
spatial and transverse momentum resolutions.
 
\subsection{Strip hit efficiencies}
\label{sec:stripEffic}

The tracklet measurement is redundant since in each RP there are five layers in
each orientation ($u$ and $v$), but three hits are enough for triggering (using
the far RPs) and for reconstruction (using all RPs). A reduced efficiency in
all detector layers in an RP, as is the case in the near bottom RPs in both
arms discussed below, results in tracklet detection inefficiency and a reduced
proton reconstruction efficiency. In these detector layers, the inefficiency is
a function of the sequential strip number in each layer and is probably caused
by two factors: the instantaneous luminosity and the radiation damage
accumulated during the lifetime of the detectors.

Strip-level efficiencies are extracted from the data by using the dominant hit
location patterns defined below. Using the library of patterns, we count the
occurrence of similar patterns where the hit position in the given layer
differs by either -2, -1.5, -1, -0.5, 0, 0.5, 1, 1.5, or 2 strip units. If the
most common pattern has a frequency of more than 90\% among the variants, and
possesses a one-strip hit in the given layer, this pattern is called dominant.

For a tracklet with no missing hit, layers are selected one by one. If the
pattern for a layer is dominant, the strip must have a hit, and the efficacy
counter ($n_1$) for this strip is increased. For a tracklet with exactly one
missing hit, the location of that hit can be predicted through the dominant
pattern, and the inefficacy counter ($n_0$) for that strip is increased. The
strip efficiency and its uncertainty are estimated using these counters.

\begin{figure}
 \centering

 \includegraphics[width=\textwidth]{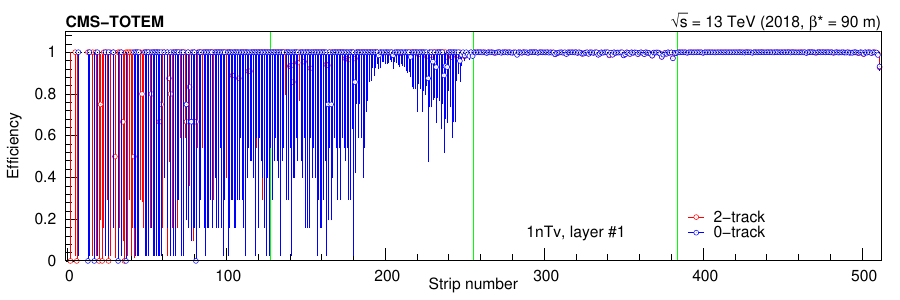}

 \includegraphics[width=\textwidth]{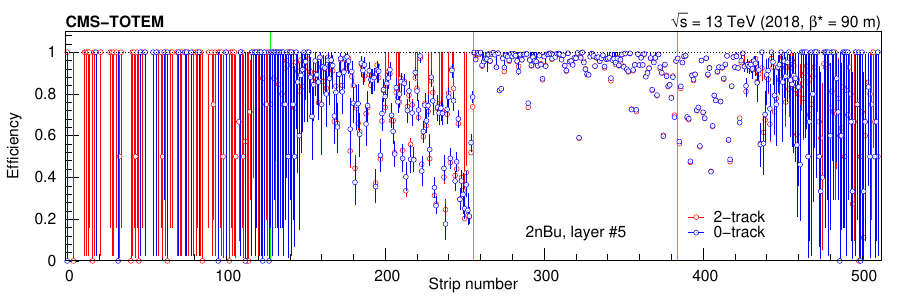}

 \caption{Strip hit efficiency from data, determined with a tag-and-probe
method. Shown here are layer 5 in layer group 1nTv (top) and 2nBu (bottom) for
a specific run. Values and statistical uncertainties based on the 2-track data
set (red symbols) and those from 0-track data (blue symbols) are plotted. The
borders of the front-end chips are indicated with (green) vertical lines. The
uncertainties are large at the left or the right sides of the layers, because
those regions are rarely hit.}

 \label{fig:stripEffic_1nTv_2nBu}

\end{figure}

\begin{figure}
 \centering
 \includegraphics[width=0.8\textwidth]{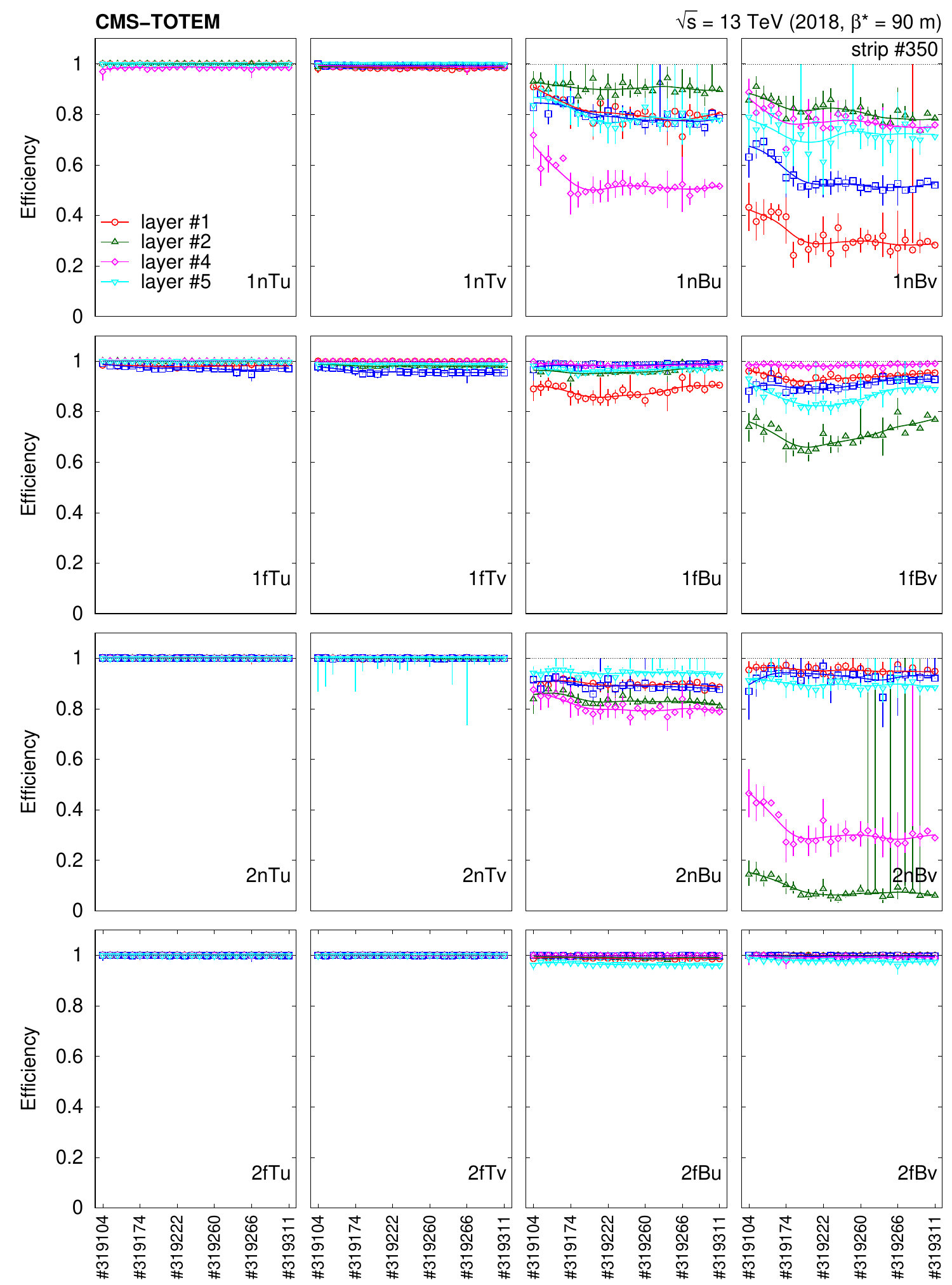}

 \caption{Strip hit efficiencies as functions of the run number for strip \#350
in layers 1--5 (coloured points) in various layer groups (labels in
the lower right corner). Statistical uncertainties are indicated with vertical
bars. Lines are spline interpolations to guide the eye.}

 \label{fig:stripEffic_vs_run}

\end{figure}

Strip hit efficiencies extracted from the data are shown for a given layer in
the 1nTv and 2nBu layer groups, and for a representative run, in
Fig.~\ref{fig:stripEffic_1nTv_2nBu}. Values based on the 2-track data set and
those from the 0-track data are plotted. They are compatible with each other,
and the effect of their uncertainty in the proton reconstruction efficiency is
about 3\%. There are regions where the efficiency is rather low--especially the
layer groups 1nBu, 1nBv, 2nBu, and 2nBv. These are the near bottom RPs on both
sides of the IP, likely the consequence of radiation damage. Strip hit
efficiencies as a function of run number, for strip \#350, as an example, in
layers 1--5 in various layer groups are shown in
Fig.~\ref{fig:stripEffic_vs_run}. The efficiencies are mostly constant, but
for some planes they change with the run number by up to 20\%.

\subsection{Tracklet efficiencies and joint weights}
\label{sec:groupEffic}

\begin{figure}[ht]
 \centering
 \includegraphics[width=0.8\textwidth]{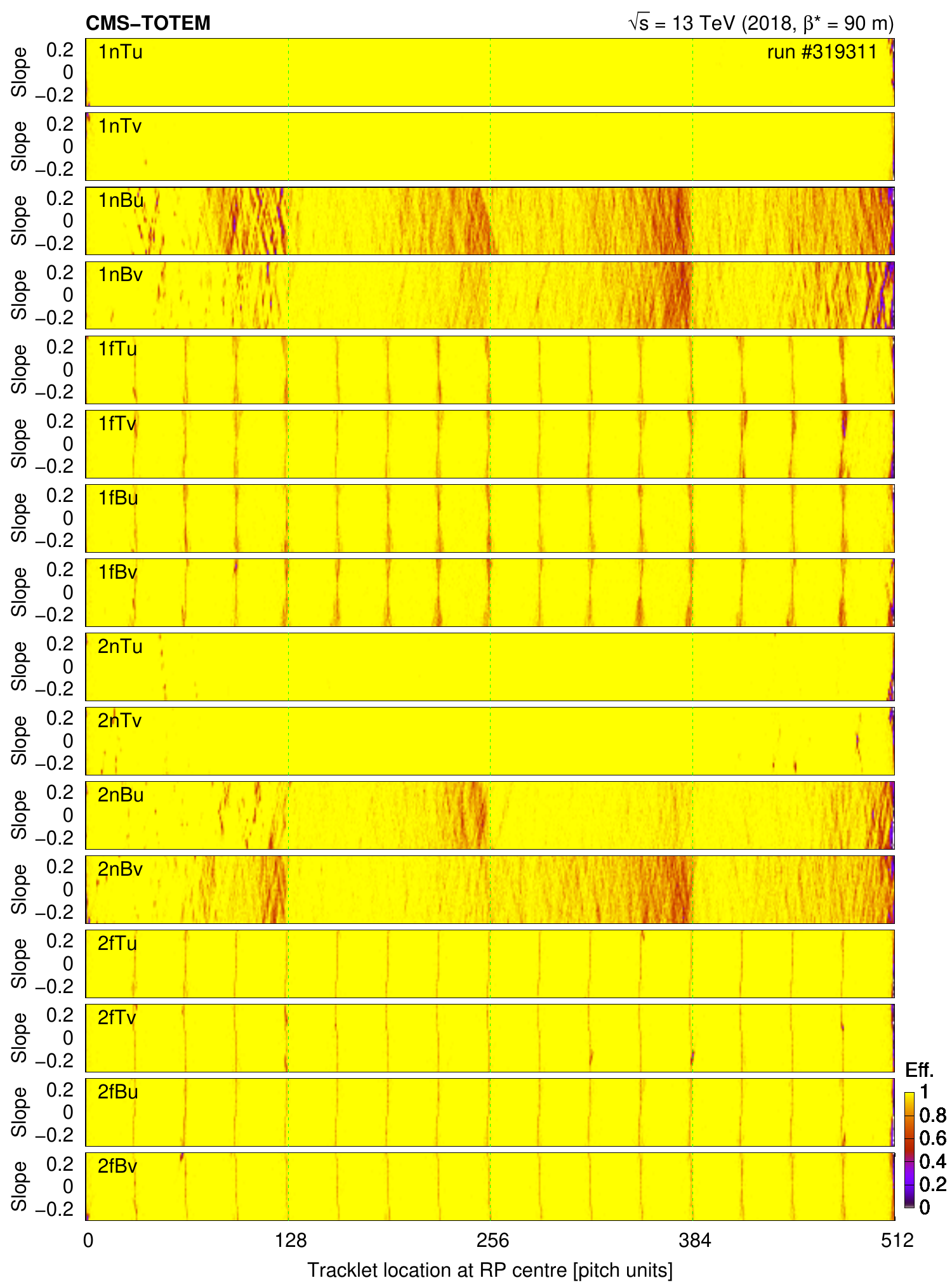}

 \caption{Tracklet reconstruction efficiency (colour scale) as a function of
tracklet location at the centre of the RP (intercept) and the track slope, for
a representative run, shown for each layer group separately. Yellow regions
correspond to fully efficient tracklet reconstruction, whereas the red regions
exhibit substantial losses with efficiencies in the range 0.4--0.6.
Vertical green lines denote boundaries of the front-end chips.}

 \label{fig:groupEffic}

\end{figure}

The tracklet reconstruction efficiencies are determined by means of a
simulation based on the strip hit detection efficiencies. Efficiency as a
function of the tracklet location at the centre of the RP (the track intercept)
and of the track slope is shown for each layer group separately in
Fig.~\ref{fig:groupEffic}, for a representative run. Substantial losses are
present at specific tracklet locations and slopes. The periodic efficiency
loss seen at every 32nd strip for the far RPs is due to the RP trigger that
uses roads of 32 strips to define a track candidate. If the tracklet is in the
border region between trigger roads, there is an increased chance that it is
lost. This kind of pattern is not observed in the near RPs that were not used
in the trigger.

In a normal event, where both scattered protons are detected by both near and
far pots, we have four tracklets, each with $u$ and $v$ projections. The
product of their corresponding tracklet efficiencies (obtained from a table,
Fig.~\ref{fig:groupEffic}) gives the probability of detecting such an event.
The reciprocal of this joint probability is the joint tracklet weight that is
applied in physics analyses. As expected from the low strip hit efficiencies in
the bottom RPs, events with trigger configuration BB suffer the most, while the
detection of TT events is fully efficient.

Once the tracklet coordinates in $u$ and $v$ directions are fitted, the hit
transverse location $(x, y)$ in the global coordinate system, at the centre of
the RP (on the ``reference surface''), is calculated through a rotation
(Section~\ref{sec:rp}).

\begin{figure}
 \centering
 \includegraphics[width=0.8\textwidth]{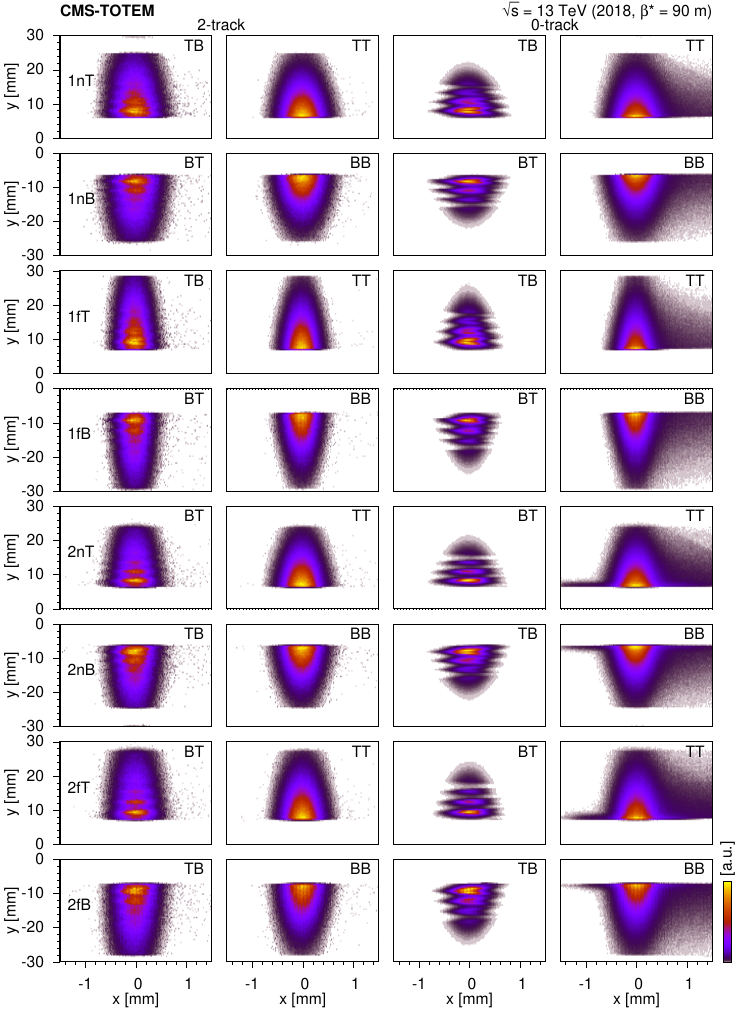}

 \caption{Efficiency-corrected distribution of proton hit locations in the
$x\text{--}y$ plane at the references surfaces of the eight RPs (rows), for
different trigger configurations (labels at the upper right corner). The two
columns on the left side refer to the 2-track data set, whereas the two on the
right side display distributions based on the 0-track data set. The wavy
pattern and the halo seen in the 0-track data set is explained in the text.}

 \label{fig:rp_location}

\end{figure}

The efficiency-corrected distribution of proton hit locations in the
$x\text{--}y$ plane at the reference surfaces of the eight RPs, are shown in
Fig.~\ref{fig:rp_location} for different trigger configurations. The wavy
pattern seen in the plots of the events with the diagonal trigger (TB and BT)
is a direct consequence of the trigger rejecting a large fraction of elastic
events. This is performed by vetoing events where hits in the two arms have
similar coordinates but with opposite sign. The halo seen in the 0-track sample
for the parallel trigger configurations (TT and BB) shows that those are not
elastic events, but overlapping inelastic events where the momentum loss of the
scattered protons is significant.

\label{sec:veto}

\clearpage

\section{Proton reconstruction}
\label{sec:proton}

Our goal is to reconstruct the $(p_x,p_y)$ transverse momentum vectors, or the
emission angles $(\theta_x^*,\theta_y^*)$, and the locations $(x^*,y^*)$ of the
scattered protons in the transverse plane at the nominal IP. This is
accomplished by using only the fitted hit locations $(x,y)$ of the proton
tracklets at the reference surfaces of the RPs. The deduced local slopes ($\rd{x}/\rd{z}$ and $\rd{y}/\rd{z}$) are not used since those have poor resolution. A
detailed cross-check of the beam optics based on the measured hit position
covariances, and a precise run-by-run alignment are also described in the
following.

\subsection{Beam optics}
\label{sec:lhc_optics}

The nominal optics determination and alignment procedure for RPs uses elastic
events~\cite{Antchev:2014voa,Nemes:2015xub} and MAD-X~\cite{Herr:2004mh}, which
is a general purpose accelerator design program. The method exploits
kinematical distributions of elastically scattered protons observed in the RPs.
It varies the rotation angles and currents of the six quadrupoles between the
IP and the RPs for both beams, as well as the beam energy of both beams, each
within its uncertainty. Theoretical predictions, as well as Monte Carlo
studies, show that the residual uncertainty in the deduced optical parameters
is smaller than 0.25\%.

The transverse coordinate $x$ of a particle (proton) at position $s$ along the
beam line can be described as~\cite{Wilson:513326}
\begin{equation}
 x(s) = \sqrt{\beta_x(s) \varepsilon}
         \; \cos\left[\phi_0 + \Delta\mu(s)\right] + D_x(s) \Delta p/p,
\end{equation}
where $\beta$ is the betatron amplitude, $\varepsilon$ is the emittance,
$\phi_0$ is the phase offset, $\Delta\mu$ is the phase advance, $D$ is the
dispersion function, and $\Delta p/p$ is the relative momentum loss of the
scattered proton. The quantities characterising the LHC optics are the
horizontal and vertical betatron amplitude $\beta(s)$, the magnification $v(s)
= \sqrt{\beta(s)/\beta^*} \cos\Delta\mu(s)$, and the effective length $L(s) =
\sqrt{\beta(s) \beta^*} \sin\Delta\mu(s)$. The phase advance $\Delta\mu(s)$ is
calculated as $\Delta\mu(s) = \int_0^s \rd{s'}/\beta(s')$, in other words $\rd{\Delta\mu}/\rd{s} = 1/\beta(s)$. At the nominal IP ($s = 0$), the location of the
particle is $x^* = x(0)$, $y^* = y(0)$, and $\beta^* = \beta(0)$, $\Delta\mu(0)
= 0$. The dependences of $x$ on $x^*$, $\theta^*$, and $\Delta p/p$ around a
given location can be linearised, and in an arm for the two RP stations
(subscript $n$ for near, and $f$ for far) we have:
\begin{align}
 x_\mathrm{n} &= v_{x,\mathrm{n}} x^* + L_{x,\mathrm{n}} \theta_x^*
               + D_{x,\mathrm{n}} \Delta p/p, \label{eq:x1} \\
 x_\mathrm{f} &= v_{x,\mathrm{f}} x^* + L_{x,\mathrm{f}} \theta_x^*
               + D_{x,\mathrm{f}} \Delta p/p, \label{eq:x2}
\end{align}
where the path-dependent dispersion function $D_x(s)$ is needed for the
description of particles with momentum spread or momentum loss ($\Delta p$).
For $\beta^* = 90\unit{m}$, the expected values are approximately
 $L_x \approx 0$, $L_y \approx 262\unit{m}$,
 $v_x \approx -1.9$, $v_y \approx 0.0$,
 $D_x \approx -0.041\unit{m}$, and $D_y \approx 0$,
based on Table~6.1 in Ref.~\cite{Niewiadomski:2008zz}. The actual values are a
bit different, asymmetric, and are shown in Table~\ref{tab:totem_optics}
(truncated to four or five significant digits), based on MAD-X calculations and
subsequent matching with data~\cite{Antchev:2014voa,Nemes:2015xub}. The optics
is such that the $x$ and the $y$ directions decouple. In the $y$ direction the
magnification and dispersion are close to zero; therefore, only the emission
angle can be determined with good precision, through a simple proportionality
relation $\theta_y^* = y_\mathrm{n}/L_{y,\mathrm{n}}$ or
$y_\mathrm{f}/L_{y,\mathrm{f}}$.

\begin{table}[bh]
 \centering
 \topcaption{Nominal values of beam optics variables (magnifications $v$,
effective lengths $L$)~\cite{Antchev:2014voa,Nemes:2015xub}, truncated to
four or five significant digits.} \label{tab:totem_optics}
 \begin{tabular}{ccrrrr}
 \hline
  Arm & Station &
  \multicolumn{1}{c}{$v_x$} & \multicolumn{1}{c}{$L_x$ [m]} &
  \multicolumn{1}{c}{$v_y$} & \multicolumn{1}{c}{$L_y$ [m]} \\
  \hline
  1 & near & $-2.204$ &  $3.1042$ &  $0.032395$ & $238.2$ \\
  1 & far  & $-1.884$ & $-0.5225$ &  $0.007509$ & $271.3$ \\
  2 & near & $-2.245$ &  $0.1943$ &  $0.018513$ & $238.3$ \\
  2 & far  & $-1.923$ & $-2.9508$ & $-0.008295$ & $271.3$ \\
 \hline
 \end{tabular}
\end{table}

For elastic and central exclusive collisions ($\abs{\Delta p/p} \ll 1$), the above
Eqs.~\eqref{eq:x1}-\eqref{eq:x2} can be solved as
\begin{align}
 x^*        &= (L_{x,\mathrm{f}} x_\mathrm{n}
              - L_{x,\mathrm{n}} x_\mathrm{f})/d, \label{eq:x-star} \\
 \theta_x^* &= (v_{x,\mathrm{n}} x_\mathrm{f}
              - v_{x,\mathrm{f}} x_\mathrm{n})/d, \label{eq:theta_x-star}
\end{align}
where the determinant is $d \equiv v_{x,\mathrm{n}} L_{x,\mathrm{f}} -
v_{x,\mathrm{f}} L_{x,\mathrm{n}}$. It can be shown that $d$ equals the
distance between the two RP stations, more precisely the distance between their
reference surfaces in the beam direction (in our case, $d = 7000\mm$).

Cross-checks of beam optics were performed with data, based on the measured
variances and covariance of the proton hit positions in $x$. Once these
quantities are measured, the variance $\var(x^*)$ is obtained as
 \begin{gather}
  \label{eq:var_xstar}
  \var(x^*) = \frac{\var(x_\mathrm{n}) \var(x_\mathrm{f}) - \cov(x_\mathrm{n},x_\mathrm{f})^2}
              {\var(x_\mathrm{f}) v_{x,\mathrm{n}}^2 - 2 \cov(x_\mathrm{n},x_\mathrm{f})
v_{x,\mathrm{n}} v_{x,\mathrm{f}} + \var(x_\mathrm{n}) v_{x,\mathrm{f}}^2}, \\
 \intertext{which does not depend on the actual values of the effective
lengths, but only on the given magnifications $v_{x,\mathrm{n}}$ and
$v_{x,\mathrm{f}}$, and the measured hit position variances and covariances. The ratio of far and near effective lengths is}
  \frac{L_{x,\mathrm{f}}}{L_{x,\mathrm{n}}} =
                    \sqrt{\frac{\var(x_\mathrm{f}) - \var(x^*) v_{x,\mathrm{f}}^2}
                               {\var(x_\mathrm{n}) - \var(x^*) v_{x,\mathrm{n}}^2}}.
 \end{gather}

 The variance of the emission angle is
 \begin{equation}
  \var(\theta_x^*) = \frac{\var(v_{x,\mathrm{f}} x_\mathrm{n} -
v_{x,\mathrm{n}}
x_\mathrm{f})} {\abs{d}^2}.
  \label{eq:det}
 \end{equation}

 With that, the effective lengths are
 \begin{align}
  L_{x,\mathrm{n}} &= \sqrt{\frac{\var(x_\mathrm{n}) - \var(x^*)
v_{x,\mathrm{n}}^2}{\var(\theta_x^*)}}, &
  L_{x,\mathrm{f}} &= \sqrt{\frac{\var(x_\mathrm{f}) - \var(x^*)
v_{x,\mathrm{f}}^2}{\var(\theta_x^*)}}.
 \end{align}

The estimated size of the beam spot in the $x$ direction is $\sigma_{x^*}
\approx 100\mum$, consistent with measurements using charged particles
reconstructed in the central CMS silicon tracker. The emission angle is
$\sigma_{\theta_x^*} \approx 28\,\mu\mathrm{rad}$ for 0-track and
$42\,\mu\mathrm{rad}$ for 2-track events. These values are substantially larger
than the beam divergence ($2.1\,\mu\text{rad}$, Section~\ref{sec:lhc_beam}), since
the characteristic width of the proton emission angle is dominated by the much
larger scale of the various physics processes. The ratio of effective lengths
for Arm 1 is $L_{x,f}/L_{x,n} \approx -0.3$, and for Arm 2 $L_{x,f}/L_{x,n}
\approx -4$. These are to be compared with the nominal values $-0.17$ and
$-15$, yielding only a poor agreement. This first look is refined, and shows a
much better agreement, once the precise run-by-run RP alignment is put in place
(Section~\ref{sec:absolute_align}).

\subsection{Effects of the IP transverse location and proton emission angle}

The correlation of the proton hit locations in the $x$ and $y$ directions
reflects various physics effects, such as momentum loss, width of the beam at
the nominal IP, and the influence of effective length ratios.

\begin{figure}[ht!]
 \centering
 \includegraphics[width=\textwidth]{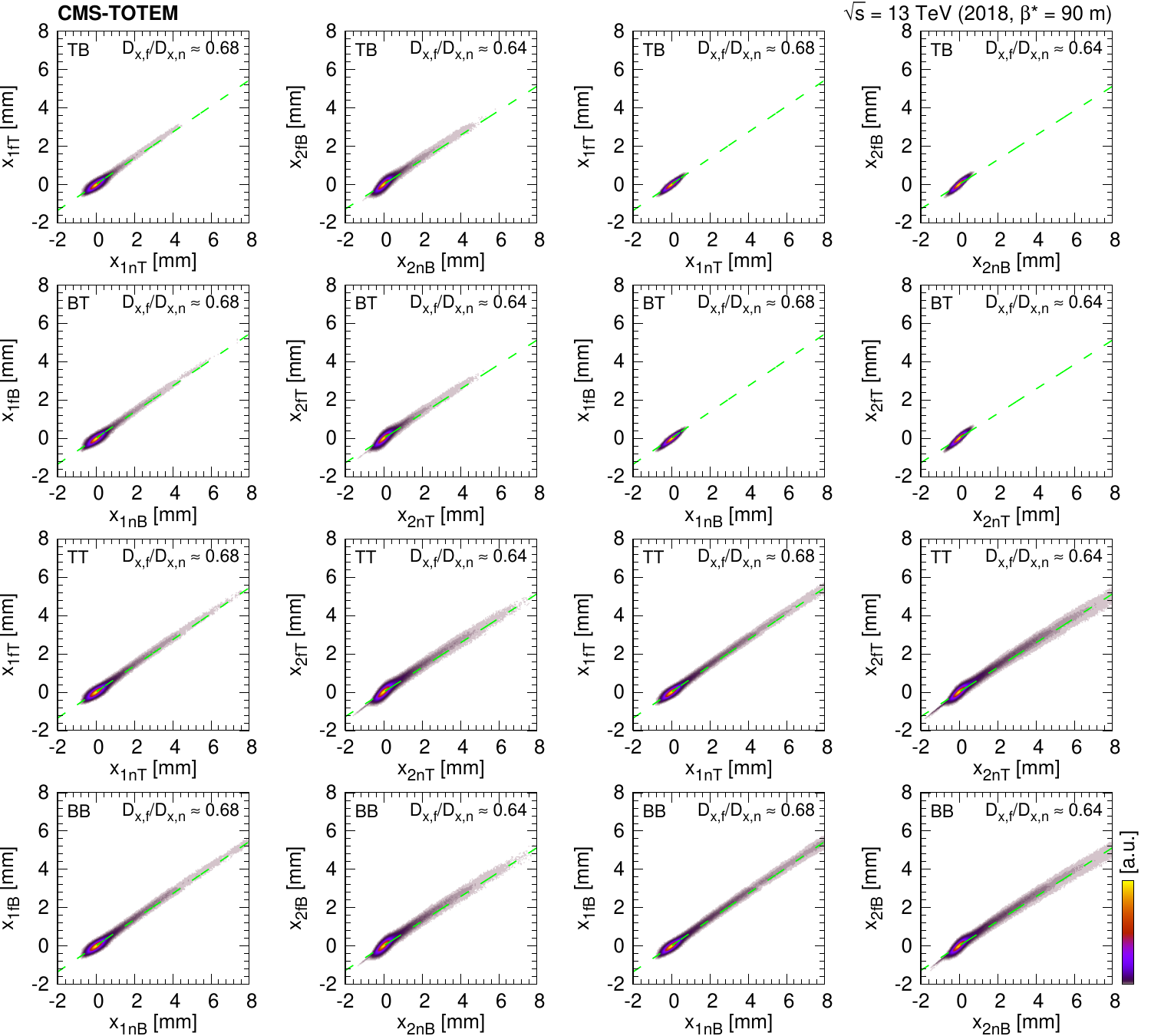}
 \caption{Correlation of proton hit locations in the $x$ direction
(two-dimensional occupancy histograms) in the far and near RPs, shown for
various trigger configurations (TB, BT, TT, and BB, in rows). The two columns
on the left side refer to the 2-track data set, and the two on the right side
display distributions based on the 0-track data set. A straight line
corresponding to the expectation $x_\mathrm{f} = x_\mathrm{n} \,
D_{x,\mathrm{f}}/D_{x,\mathrm{n}}$ is also plotted, where $D_{x,f}/D_{x,n}
\approx 0.68$ for Arm 1, and $\approx 0.64$ for Arm 2. The plots are produced
with the final detector alignment.}
 \label{fig:rp_nf_x}
\end{figure}

For elastic collisions, the momentum loss of the protons is very small, its
spread is $\abs{\Delta p/p} \lesssim 10^{-4}$, reflecting the momentum variance of
the beam. This corresponds to about $5\mum$ hit position spreading, which is
hardly detectable. The acceptance of the central tracker limits the detectable
2-track central exclusive events to $\abs{\Delta p/p} \lesssim 10^{-3}$ through
momentum conservation, corresponding to a $50\mum$ shift in hit position
(comparable to the strip width of $66\mum$). Here ``shift'' refers to the
difference between the true hit position and the predicted track impact point
position assuming the nominal beam momentum. For diffractive and highly
inelastic collisions, where the momentum loss of the incoming protons is
sizeable, a well-detectable displacement is visible: a 10\% momentum loss
corresponds to a $5\mm$ shift. The events show a linear correlation between
$x_\mathrm{f}$ and $x_\mathrm{n}$ (Fig.~\ref{fig:rp_nf_x}) with a slope
$D_{x,f}/D_{x,n} \approx 0.68$ for Arm 1, and $\approx 0.64$ for Arm 2, as
expected.

The effect of the beam width at the nominal IP, $\sigma_{x,y}^* \sim {\cal
O}(100\mum)$, can be propagated to the RPs ($v_x x^*$) and results in a spread
of about $200\mum$. Such an event-by-event shift is correlated between the two
arms (the protons come from the same actual IP) and is well detectable. For a
transverse momentum $p_x = 200\,\mathrm{MeV}$ we have $\theta_x^* \approx
10^{-4}$, which results in a shift of $20\text{--}300\mum$. The higher values
are already detectable and in the case of elastic collisions there should be a
clear correlation between the two arms.

\begin{figure}
 \centering
 \includegraphics[width=\textwidth]{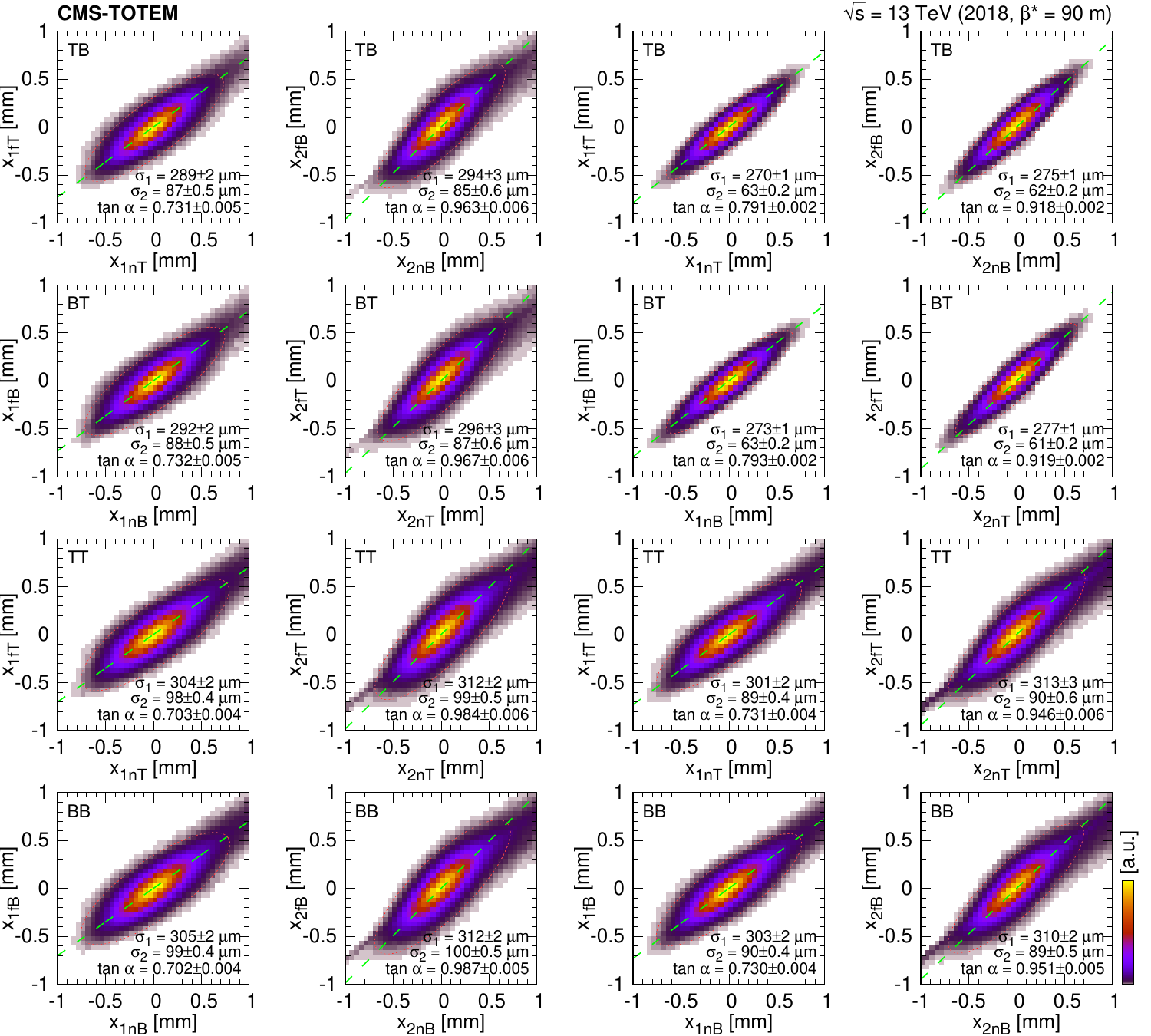}
 \caption{Correlation of proton hit locations in the $x$ direction
(two-dimensional occupancy histograms) in the far and near RPs, shown for
various trigger configurations (TB, BT, TT, and BB, in rows), with a restricted
scale. Parameters (standard deviations in major and minor axis directions
$\sigma_1$ and $\sigma_2$, and the rotation angle $\alpha$) of the fitted
tilted two-dimensional normal distributions are displayed in the plots. The
corresponding ellipses cover 99\% of the points; the dashed green lines
indicate their major axes. The two columns on the left side refer to the
2-track data set, whereas the two on the right side display distributions based
on the 0-track data set. The plots are produced with the final detector
alignment.}
\label{fig:rp_nf_x_zoom}
\end{figure}

In the case of small $x$ values (Fig.~\ref{fig:rp_nf_x_zoom}), the distribution
consists mostly of elastic and central exclusive events with $\Delta p/p
\approx 0$. For the diagonal configuration, the 0-track data set (mostly
elastic events) shows a slope of 0.79 for Arm 1, and 0.92 for Arm 2. For the
parallel configuration (in part central exclusive events), the slope is
0.70--0.73 for Arm 1, and 0.95--0.99 for Arm 2. The slopes
for elastic and central exclusive events differ because the variance of the
proton emission angle $\var(\theta_x^*)$ in the two cases is different. The
quantities shown in Fig.~\ref{fig:rp_nf_x_zoom} are used for estimating
$\var(x^*)$ using Eq.~\eqref{eq:var_xstar}.

\begin{figure}
\centering
\includegraphics[width=\textwidth]{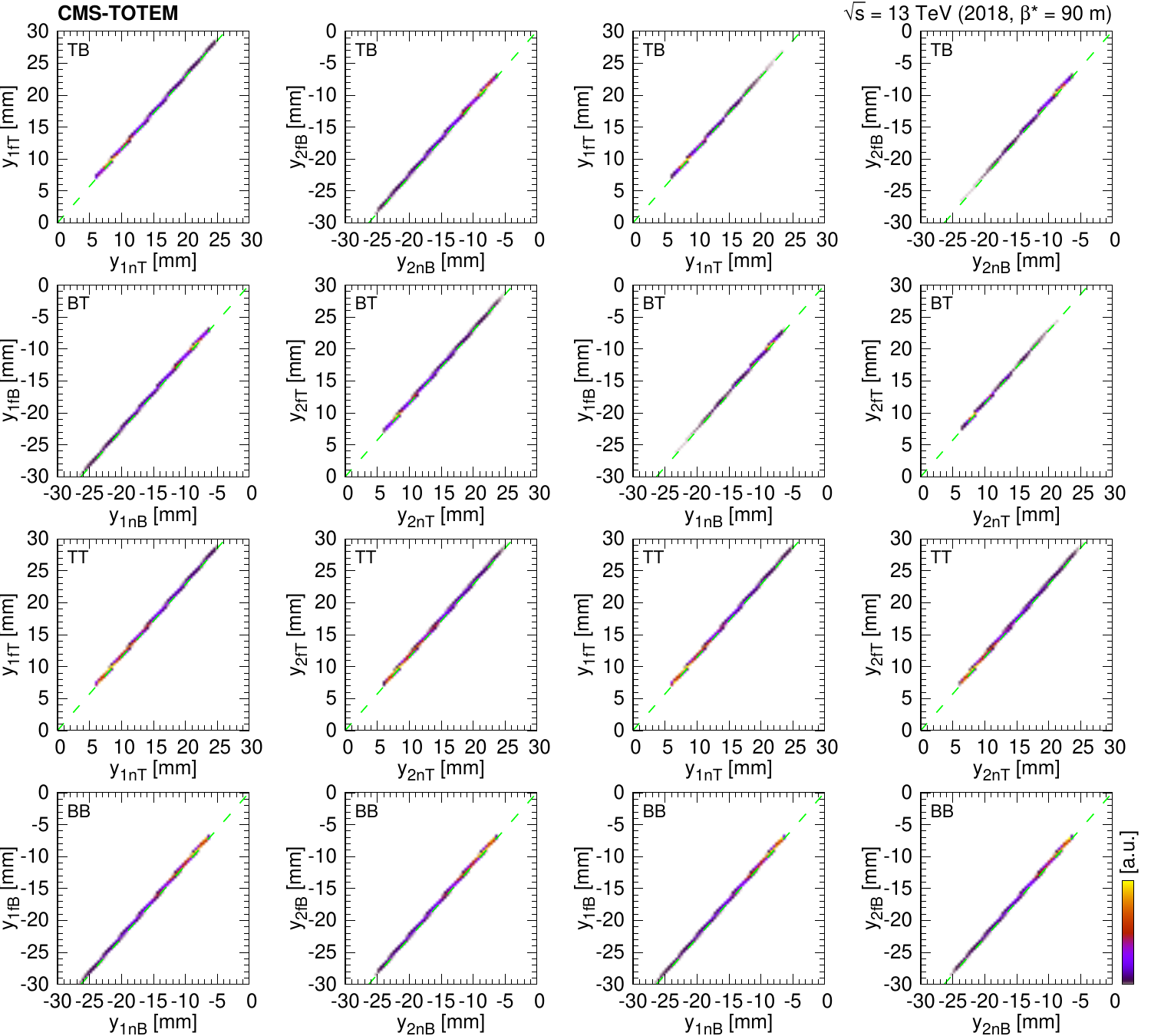}

 \caption{Correlation of proton hit locations in the $y$ direction
(two-dimensional occupancy histograms) in the far and near RPs, shown for
various trigger configurations (TB, BT, TT, and BB, in rows). The two columns
on the left side refer to the 2-track data set, whereas the two on the right
side display distributions based on the 0-track data set. A straight line
corresponding to the expectation $y_\mathrm{f} = y_\mathrm{n} \,
L_{y,\mathrm{f}}/L_{y,\mathrm{n}}$ is also plotted, where $L_{y,f}/L_{y,n}
\approx 1.14$. The plots are produced with the final detector alignment. The
apparent piecewise linear segments are simply the consequence of the binning.}
 \label{fig:rp_nf_y}
\end{figure}

{\tolerance=800
In the $y$ direction, the measured data match our expectations with the slope
$L_{y,f}/L_{y,n} = 271.3 / 238.2 \approx 1.14$ (Fig.~\ref{fig:rp_nf_y}).
\par}


\subsection{Absolute run-by-run alignment}
\label{sec:absolute_align}

An inspection of collision events shows apparent deviations from momentum
conservation. This is corrected by means of additional shifts of the RPs in the
$x$ and $y$ directions. This absolute alignment is tested through symmetry and
momentum conservation by studying the distributions of the deduced IP
coordinates and of the momentum sum, and more specifically their means.

The alignment in both the $x$ and $y$ directions must be performed for each run
separately, since sizeable apparent displacements from run to run are observed.
The changes are mostly due to the drifting LHC proton orbit, since the RP
movement system ensures a position reproducibility of about 20\mum. However,
instead of correcting the incoming proton direction, for the sake of
simplicity, in the present analysis the RPs are artificially shifted. The goal
is to determine the additional shifts $\delta x$ and $\delta y$ for the eight
RPs, \ie, altogether 16 values.
 
The measured deviations $\vec{\Delta}$ and the local shifts $\vec{x_0}$ are
connected through a linear transformation $A \vec{x_0} = \vec{\Delta}$, where
$A$ is a matrix. The shifts $\vec{\delta x}$ are such that $A (\vec{x_0} +
\vec{\delta x}) = 0$, and therefore they can be determined from solving $A
\vec{\delta x} = -\vec{\Delta}$.

The alignment in the $x$ direction is based on the nearly Gaussian hit location
and momentum sum distributions, since they must be symmetric about zero:

\begin{itemize}

 \item the location of the primary interaction $x^* = (x_\mathrm{n}
L_{x,\mathrm{f}} - x_\mathrm{f} L_{x,\mathrm{n}})/d$. This quantity is
calculated in a single arm (1 or 2), using either the top or the bottom pots (T
or B), hence the notations 1T, 1B, 2T, and 2B.

 \item the sum of particle momenta $\sum p_x$ in elastic or exclusive events.
For scattered protons in the RPs, $p_x$ is calculated as $p_x = p_\text{beam}
(x_\mathrm{f} v_{x,\mathrm{n}}- x_\mathrm{n} v_{x,\mathrm{f}})/d$.

\end{itemize}

We can use the measured means of 16 distributions: the location of the IPs
$x^*$, calculated from RP arm configurations 1T, 1B, 2T, and 2B (4 values); the
sum of the proton transverse momenta $\sum p_x$ for RP trigger configurations
TB, BT, TT, and BB (4 values); the local hit positions $x$ in each RP (8
values). The transformation matrix is
\begin{equation}
 A_x = 
  \begin{pmatrix}
  L_\text{1f}/d & - L_\text{1n}/d & 0 & 0 & 0 & 0 & 0 & 0 \\
  0 & 0 & L_\text{1f}/d & - L_\text{1n}/d & 0 & 0 & 0 & 0 \\
  0 & 0 & 0 & 0 & L_\text{2f}/d & - L_\text{2n}/d & 0 & 0 \\
  0 & 0 & 0 & 0 & 0 & 0 & L_\text{2f}/d & - L_\text{2n}/d \\
  - pv_\text{1f}/d & pv_\text{1n}/d & 0 & 0 & 0 & 0 & -pv_\text{2f}/d & pv_\text{2n}/d \\
  0 & 0 & - pv_\text{1f}/d & pv_{1n}/d & -pv_\text{2f}/d & pv_\text{2n}/d & 0 & 0 \\
  - pv_\text{1f}/d & pv_\text{1n}/d & 0 & 0 & -pv_\text{2f}/d & pv_{2n}/d & 0 & 0 \\
  0 & 0 & - pv_\text{1f}/d & pv_\text{1n}/d & 0 & 0 & -pv_\text{2f}/d & pv_\text{2n}/d
  \end{pmatrix},
\end{equation}
where $p = p_\text{beam}$, and the transformation itself is
\begin{equation}
 A_x 
 \begin{pmatrix}
  \delta x_\text{1nT} \\
  \delta x_\text{1fT} \\
  \delta x_\text{1nB} \\
  \delta x_\text{1fB} \\
  \delta x_\text{2nT} \\
  \delta x_\text{2fT} \\
  \delta x_\text{2nB} \\
  \delta x_\text{2fB}
 \end{pmatrix} =
 \begin{pmatrix}
  - \langle x^*_\text{1T} \rangle \\
  - \langle x^*_\text{1B} \rangle \\
  - \langle x^*_\text{2T} \rangle \\
  - \langle x^*_\text{2B} \rangle \\
  - \langle \sum p_{x,\text{TB}} \rangle \\
  - \langle \sum p_{x,\text{BT}} \rangle \\
  - \langle \sum p_{x,\text{TT}} \rangle \\
  - \langle \sum p_{x,\text{BB}} \rangle
 \end{pmatrix}.
\end{equation}

\begin{figure}[th]
 \centering
 \includegraphics[width=0.8\textwidth]{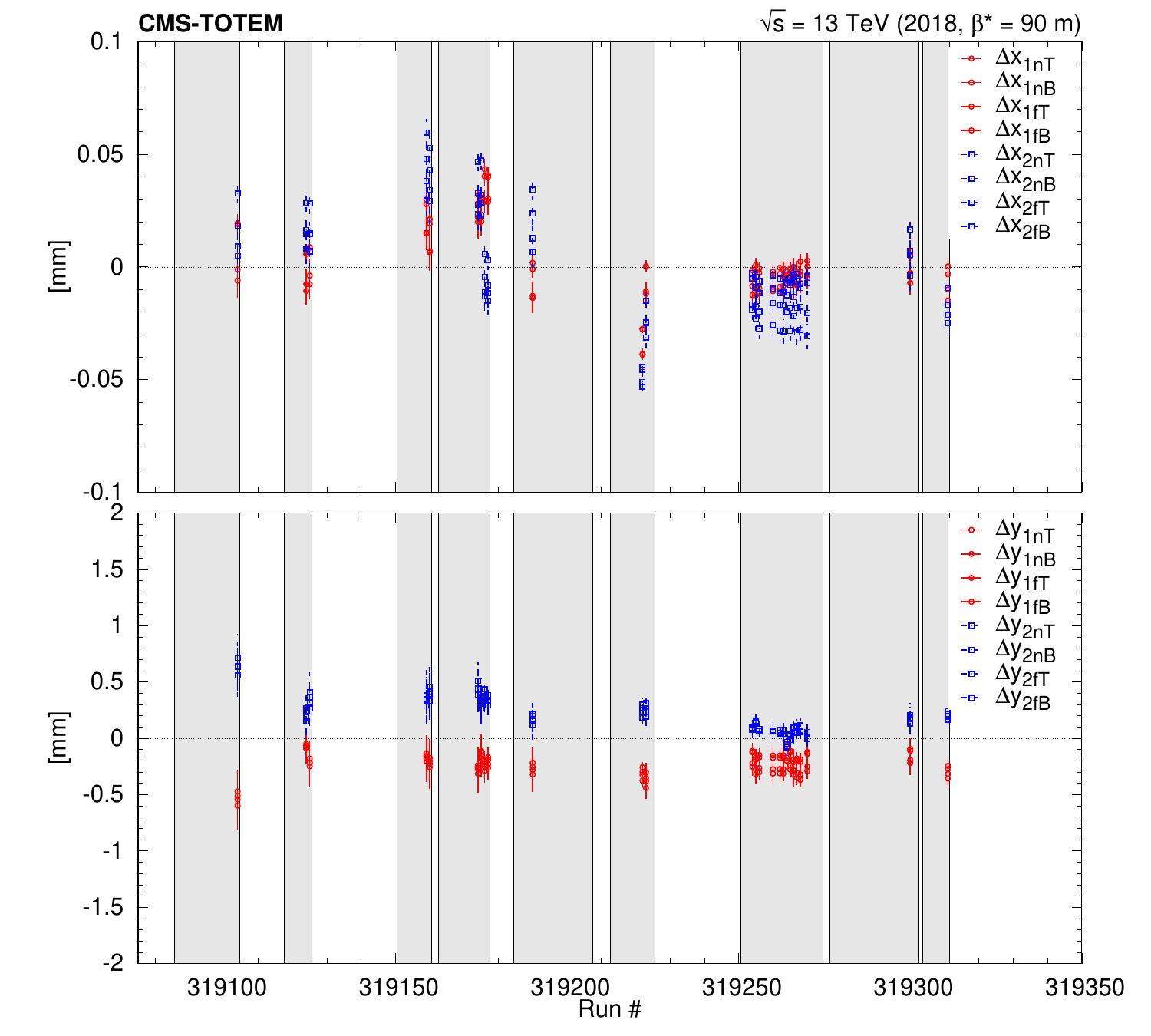}

 \caption{Deduced displacements of RPs in the $x$ (top) and $y$ (bottom)
directions as a function of run number. Statistical uncertainties are
shown with vertical bars. The LHC fills are indicated by grey areas.}

 \label{fig:org_align_first}

\end{figure}

The top and bottom RPs are independently movable, and their hit location
distribution in the $y$ direction has a gap in the middle. In addition, because
of the varying $p_y$ acceptance of various pots, the normalisation of the
distributions in the top and bottom parts is unrelated. Hence the alignment in
the $y$ direction cannot be based on the hit location distributions alone.
There are two variables whose distributions must be symmetric about zero, and
are not biased:

\begin{itemize}

 \item the location of the primary interaction $y^*$. In an arm, using either
the top or the bottom RPs, it is calculated as
 $y^* = (y_\mathrm{n} L_{y,\mathrm{f}} - y_\mathrm{f} L_{y,\mathrm{n}})/d$.
 There are four such independent equations (two arms, top or bottom RPs).

 \item the sum of particle momenta $\sum p_y$ in elastic or exclusive events.
For scattered protons in the RPs, $p_y$ is calculated as
 $p_y = p_\text{beam} [2 (y_\mathrm{n}/L_{y,\mathrm{n}}
                        + y_\mathrm{f}/L_{y,\mathrm{f}})]$.
 There are four such independent equations, one for each configuration.

\end{itemize}

We have the measured means of twelve distributions: the location of the IPs
$y^*$, calculated from RP pairs 1T, 1B, 2T, and 2B (4 values); the sum of
proton transverse momenta $\sum p_y$ for RP trigger configurations TB, BT, TT,
and BB (4 values); the averaged local hit positions $y$ in RP groups 1n, 1f,
2n, and 2f (4 values). The transformation matrix is
\begin{equation}
 A_y =
  \begin{pmatrix}
   L_\text{1f}/d & - L_\text{1n}/d & 0 & 0 & 0 & 0 & 0 & 0 \\ 
   0 & 0 & L_\text{1f}/d & - L_\text{1n}/d & 0 & 0 & 0 & 0 \\
   0 & 0 & 0 & 0 & L_\text{2f}/d & - L_\text{2n}/d & 0 & 0 \\
   0 & 0 & 0 & 0 & 0 & 0 & L_\text{2f}/d & - L_\text{2n}/d \\
    p/(2L_\text{1n}) & p/(2L_\text{1f}) & 0 & 0 & 0 & 0 & p/(2L_\text{2n}) & p/(2L_\text{2f}) \\
   0 & 0 &  p/(2L_\text{1n}) & p/(2L_\text{1f}) & p/(2L_\text{2n}) & p/(2L_\text{2f}) & 0 & 0 \\
    p/(2L_\text{1n}) & p/(2L_\text{1f}) & & & p/(2L_\text{2n}) & p/(2L_\text{2f}) & 0 & 0 \\
   0 & 0 & p/(2L_\text{1n}) & p/(2L_\text{1f}) & 0 & 0 & p/(2L_\text{2n}) &
p/(2L_\text{2f})
   \end{pmatrix},
  \end{equation}
and the transformation itself is
\begin{equation}
 A_y
 \begin{pmatrix}
  \delta y_\text{1nT} \\
  \delta y_\text{1fT} \\
  \delta y_\text{1nB} \\
  \delta y_\text{1fB} \\
  \delta y_\text{2nT} \\
  \delta y_\text{2fT} \\
  \delta y_\text{2nB} \\
  \delta y_\text{2fB}
 \end{pmatrix} =
 \begin{pmatrix}
  - \langle y^*_\text{1T}\rangle \\
  - \langle y^*_\text{1B}\rangle \\
  - \langle y^*_\text{2T}\rangle \\
  - \langle y^*_\text{2B}\rangle \\
  - \langle \sum p_{y,\text{TB}} \rangle \\
  - \langle \sum p_{y,\text{BT}} \rangle \\
  - \langle \sum p_{y,\text{TT}} \rangle \\
  - \langle \sum p_{y,\text{BB}} \rangle
 \end{pmatrix}.
\end{equation}

The optimisation of the additional shifts was performed through a joint least
squares fit using the estimated variances of $\langle x^* \rangle$ and
$\langle y^* \rangle$ values, and those of the momentum sums. The extracted
values in the $x$ and $y$ directions as a function of run number are shown in
Fig.~\ref{fig:org_align_first}. All RPs in an arm need to be artificially
shifted in a coordinated manner, although the two arms appear to be
independent. The necessary run-by-run adjustments are in the range $\pm
50\mum$ in $x$, and $\pm 0.5\mm$ in $y$. The time dependence of the
additional shifts was examined, but they show no obvious periodic behaviour.
We conclude that the observed behaviour does not reflect an actual
displacement of the RPs, but the drifting of the incoming proton beam orbits.
The difference between the $x$ and $y$ directions is the consequence of the
special LHC beam optics configuration at the location of the RPs
(Section~\ref{sec:lhc_optics}). The effective length in the $x$ direction is
close to zero ($L_x \approx 0$), hence the horizontal position of the proton
in the RP principally depends on the IP coordinate $x^*$. The magnification
in the $y$ direction is close to zero ($v_y \approx 0$) which makes the
vertical position of the proton in the RP mainly dependent on the scattering
angle $\theta_y^*$ and not the IP coordinate $y^*$. Hence any beam-related
effect plays out rather differently in the two directions.

\begin{figure}

 \centering
 \includegraphics[width=0.8\textwidth]{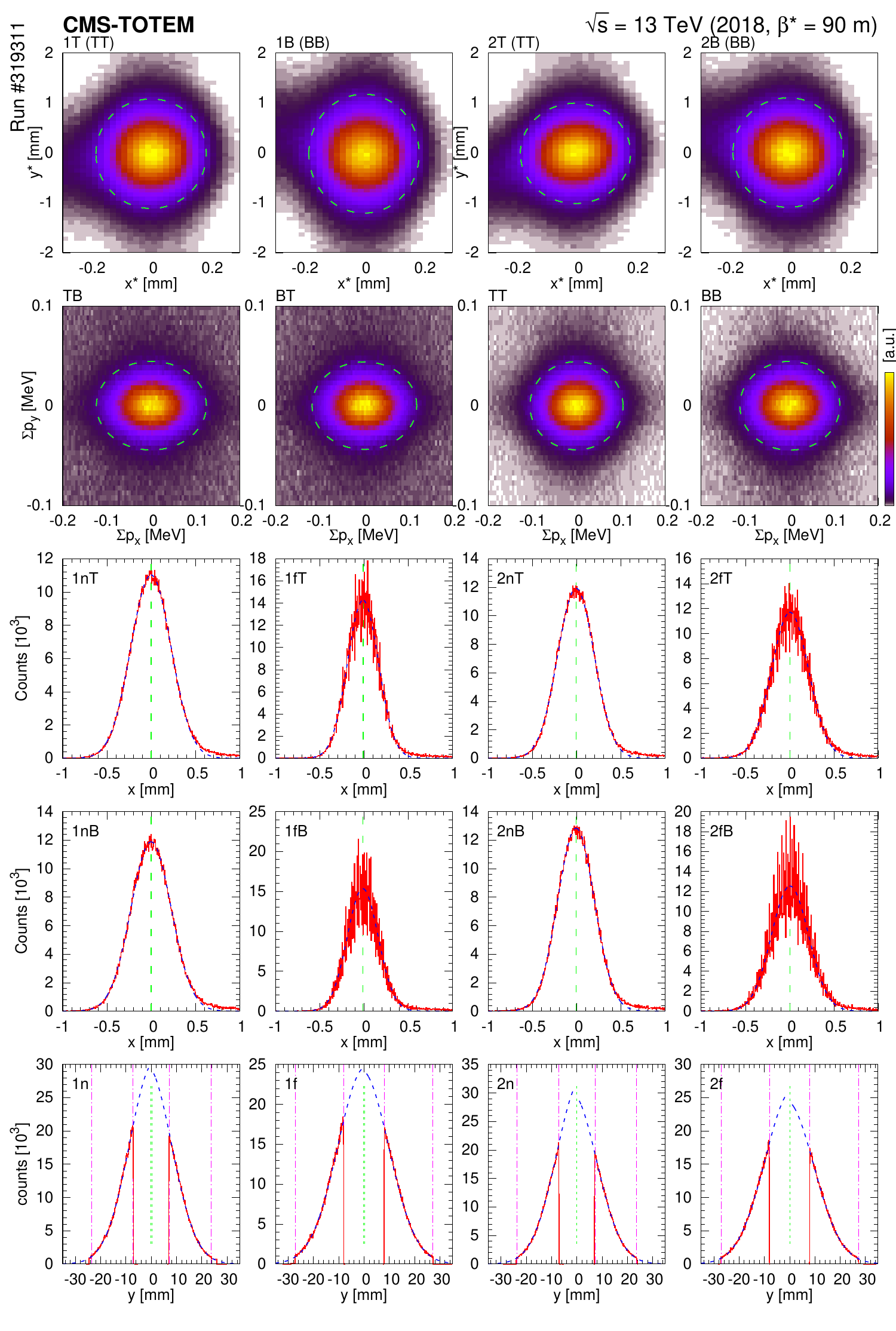}

 \caption{Cross-check of the full alignment, shown here for run \#319311. From
top to bottom: Distribution of the IP coordinates $(x^*,y^*)$ for RP arm
configurations 1T, 1B, 2T, and 2B. Distribution of the four-particle momentum
sum $(\sum p_x,\sum p_y)$ for the RP trigger configurations TB, BT, TT, and BB.
In both cases the two-dimensional Gaussian fits are indicated (at $2\sigma$)
with green dotted ellipses. Distribution of local hits in the RPs in the $x$
(single Gaussian) and $y$ directions (separate Gaussians). Dashed blue curves
represent the Gaussian fits, vertical green dashed lines indicate the deduced
relative shifts, vertical magenta dash-dotted lines on $y$ plots show fit
ranges.}

 \label{fig:upd_fit_319311_first}

\end{figure}

\begin{figure}
 \centering
 \includegraphics[width=\textwidth]{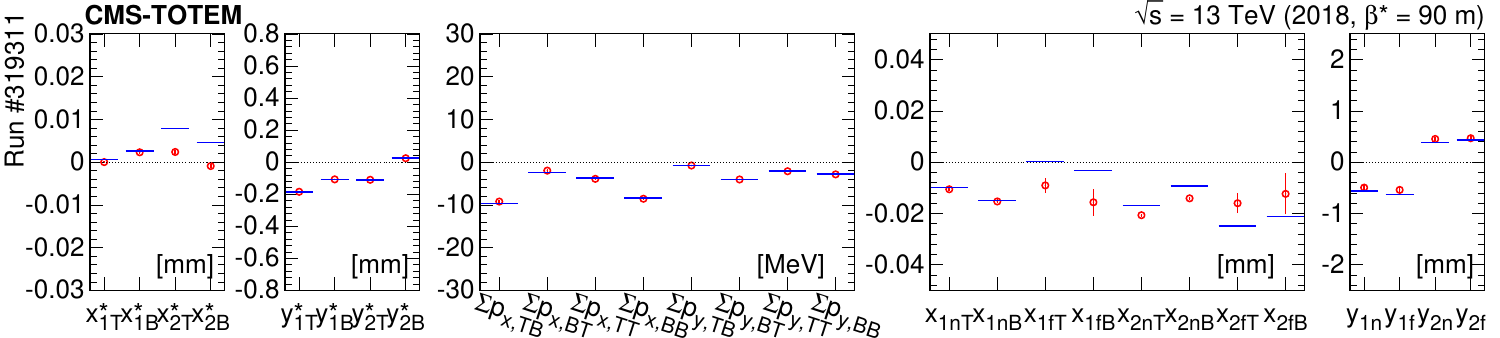}
 \includegraphics[width=\textwidth]{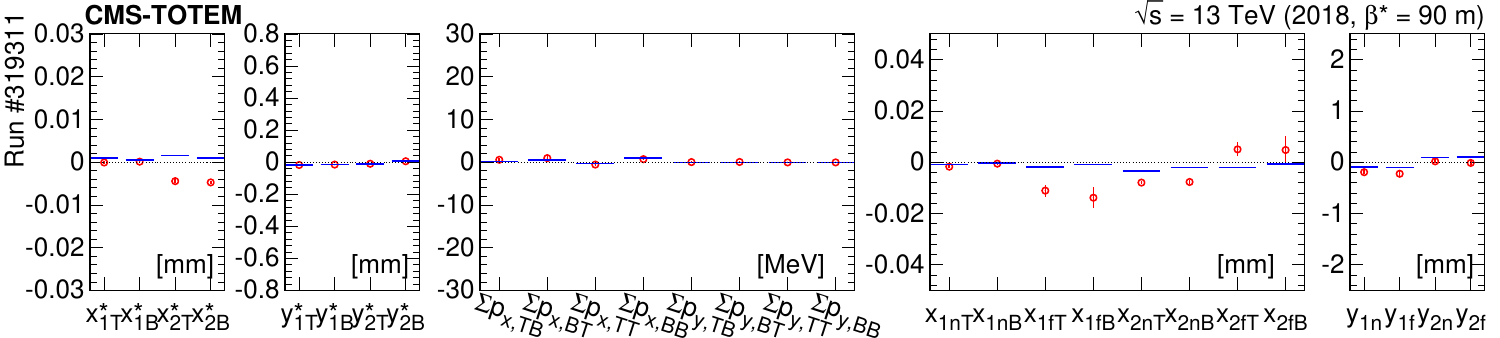}
 \caption{The measured residuals (red symbols) and those expected from the
extracted displacements (horizontal blue lines) for run \#319311, before
(upper) and after (lower) the alignment.}

 \label{fig:upd_align_319311}

\end{figure}

Details and cross-checks for a specific run (\#319311) after full alignment are
shown in Fig.~\ref{fig:upd_fit_319311_first}, including the distribution of the
IP coordinates $(x^*,y^*)$ and that of the four-particle momentum sum $(\sum
p_x,\sum p_y)$ for each RP configuration, as well as the distribution of the
local hit coordinates in the $x$ (single Gaussian) and $y$ directions (separate
Gaussians). The measured residuals and those expected from the extracted
displacements are shown in Fig.~\ref{fig:upd_align_319311}. Although not all
the constraints can be satisfied at the same time, the relevant quantities used
for event selection and physics analysis, \ie, the momentum sums $\sum p_x$ and
$\sum p_y$, are well optimised.

\begin{figure}[ht]
 \centering
 \includegraphics[width=\textwidth]{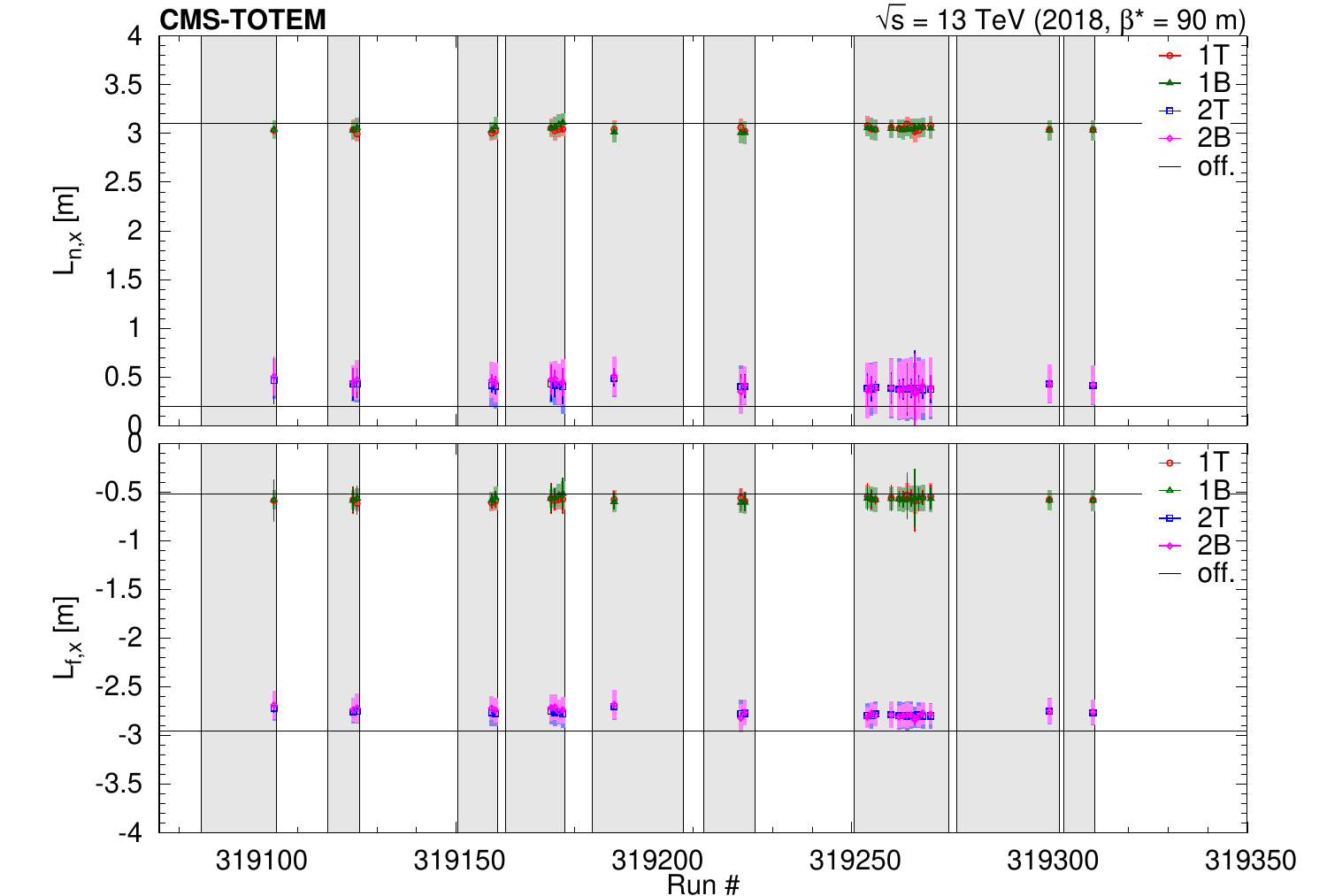}
 \caption{Effective lengths at the location of the near ($L_{n,x}$) and far
pots ($L_{f,x}$) as functions of the run number, deduced from near-far hit
covariances in RPs for RP arm configurations 1T, 1B, 2T, and 2B. Statistical
uncertainties are indicated with error bars, systematic ones are plotted with
shaded rectangles. Values of the nominal TOTEM optics parameters
(Table~\ref{tab:cov_L}) are also shown with horizontal black lines. LHC fills
are indicated by grey areas.}
 \label{fig:upd_cov_L}
\end{figure}

Estimates, based on measured data, of the effective lengths at the location of
the near ($L_{n,x}$) and far pots ($L_{f,x}$) as functions of the run number
are shown in Fig.~\ref{fig:upd_cov_L}. They were deduced from near-far hit
covariances in RPs for arm configurations 1T, 1B, 2T, and 2B. Systematic
uncertainties are obtained by varying the size of the selection ellipse
(Fig.~\ref{fig:rp_nf_x_zoom}) by 50\%, and by varying the values of
magnification ($v_{x,\text{n}}$ and $v_{x,\text{f}}$) within their expected
uncertainties (0.2\%). Values of the nominal
optics~\cite{Antchev:2014voa,Nemes:2015xub} are also plotted for reference,
with black lines. The comparison between the nominal and averaged estimates
from collected events is given in Table~\ref{tab:cov_L} where any run-by-run
variation of the estimated value is included in the uncertainty. They are in
good agreement.

\begin{table}[bh]
\centering
\topcaption{Nominal values and estimates of the effective lengths
$L_x$, truncated to two decimal places. For the estimates, the systematic
uncertainties are indicated; the statistical ones are negligible.}
 \label{tab:cov_L}
 \begin{tabular}{ccrr}
 \hline
  \multirow{2}{*}{Arm} &
  \multirow{2}{*}{Station} & \multicolumn{2}{c}{$L_x$ [m]} \\
  & & \multicolumn{1}{c}{Nominal}
    & \multicolumn{1}{c}{Estimated} \\
 \hline
  1 & near & $ 3.10$ & $ 3.05 \pm 0.11$ \\
  1 & far  & $-0.52$ & $-0.57 \pm 0.12$ \\
  2 & near & $ 0.19$ & $ 0.39 \pm 0.16$ \\
  2 & far  & $-2.95$ & $-2.78 \pm 0.12$ \\
 \hline
 \end{tabular}
\end{table}

\clearpage

\subsection{Results}
\label{sec:results}

\begin{figure}[ht]
 \centering
 \includegraphics[width=\textwidth]{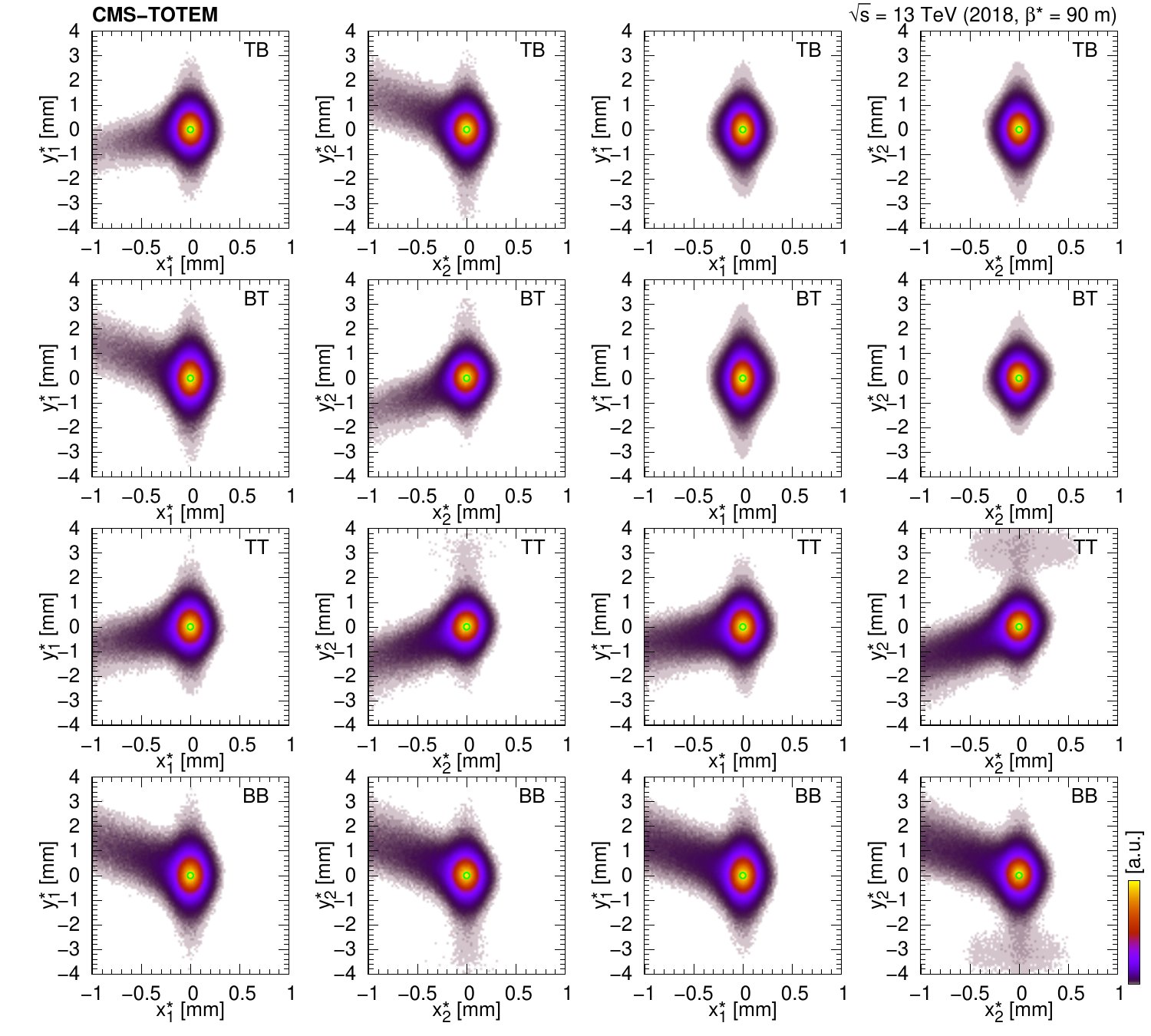}
 \caption{Location of the primary pp interaction in the $x^*\text{--}y^*$ plane
at the IP using RPs in Arm 1 or 2 (subscripts 1 or 2), shown for various
trigger configurations (TB, BT, TT, and BB, in rows). The two columns on the
left side refer to the 2-track data set, whereas the two on the right side
display distributions based on the 0-track data set. The elongated regions at
negative $x^*$ values correspond to events with large proton momentum loss.
The green circles mark $(0,0)$.}
 \label{fig:proton_loc}
\end{figure}

\begin{figure}[th]
 \centering
 \includegraphics[width=\textwidth]{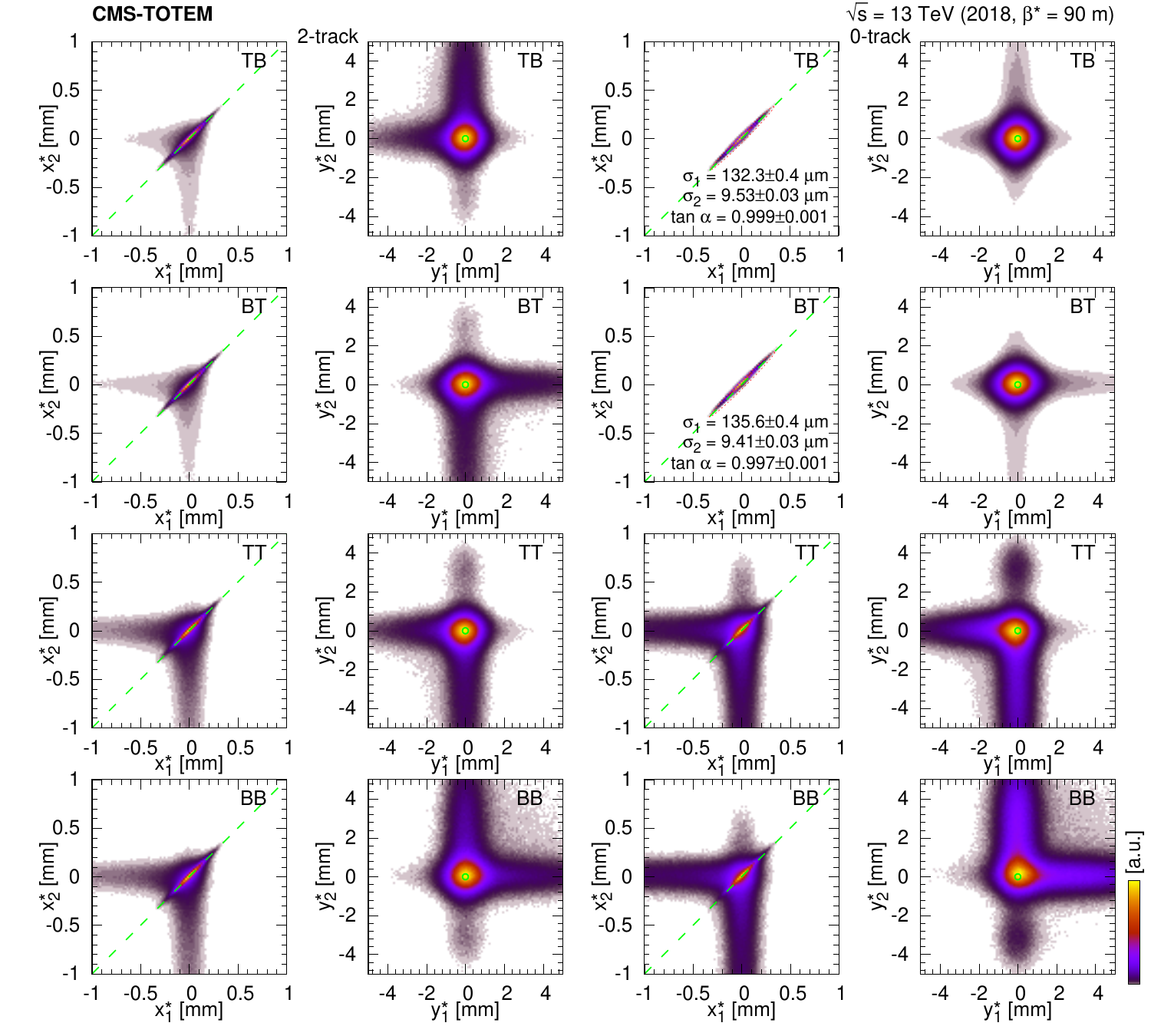}
 \caption{Joint distribution of $x^*$ (or $y^*$) coordinates deduced using RPs
in Arm 1 and 2 (subscripts 1 and 2), shown for various trigger configurations
(TB, BT, TT, and BB, in rows). The two columns on the left side refer to the
2-track data set, whereas the two on the right side display distributions based
on the 0-track data set. In the case of the diagonally triggered (TB and BT)
0-track (in part elastic) events the parameters (standard deviations in major
and minor axis directions $\sigma_1$ and $\sigma_2$, and the rotation angle
$\alpha$) of the fitted ellipses are displayed in the plots. The green circles
mark $(0,0)$, the dashed green lines are the identity lines.}\label{fig:vtx}
\end{figure}

The locations of the primary pp interactions in the $x^*\text{--}y^*$ plane at
the IP using RPs in Arm 1 or in Arm 2 are shown for each trigger configuration
in Fig.~\ref{fig:proton_loc}. All the distributions are well centred at
$(0,0)$. The elongated regions at negative $x^*$ values correspond to protons
with large momentum loss. The width of the distributions is the result of
various factors: the beam spot size, the effect of multiple scattering of the
protons within the RP material, and the local position resolution of the RPs.
The joint distributions of $x^*$ (or $y^*$) coordinates deduced using RPs in
Arm 1 and 2 are shown for each trigger configuration in Fig.~\ref{fig:vtx}. The
$x^*$ coordinates are determined with good precision, and they correlate well
between Arm 1 and 2, with a beam spot size of $135/\sqrt{2} \approx 95\mum$ and
resolution $10/\sqrt{2} \approx 6\text{--}7\mum$. The $y^*$ coordinates have
much larger uncertainties, but their distributions are well centred around
$(0,0)$, with tails from non-exclusive events.

\clearpage

\begin{figure}[th]
 \centering
 \includegraphics[width=\textwidth]{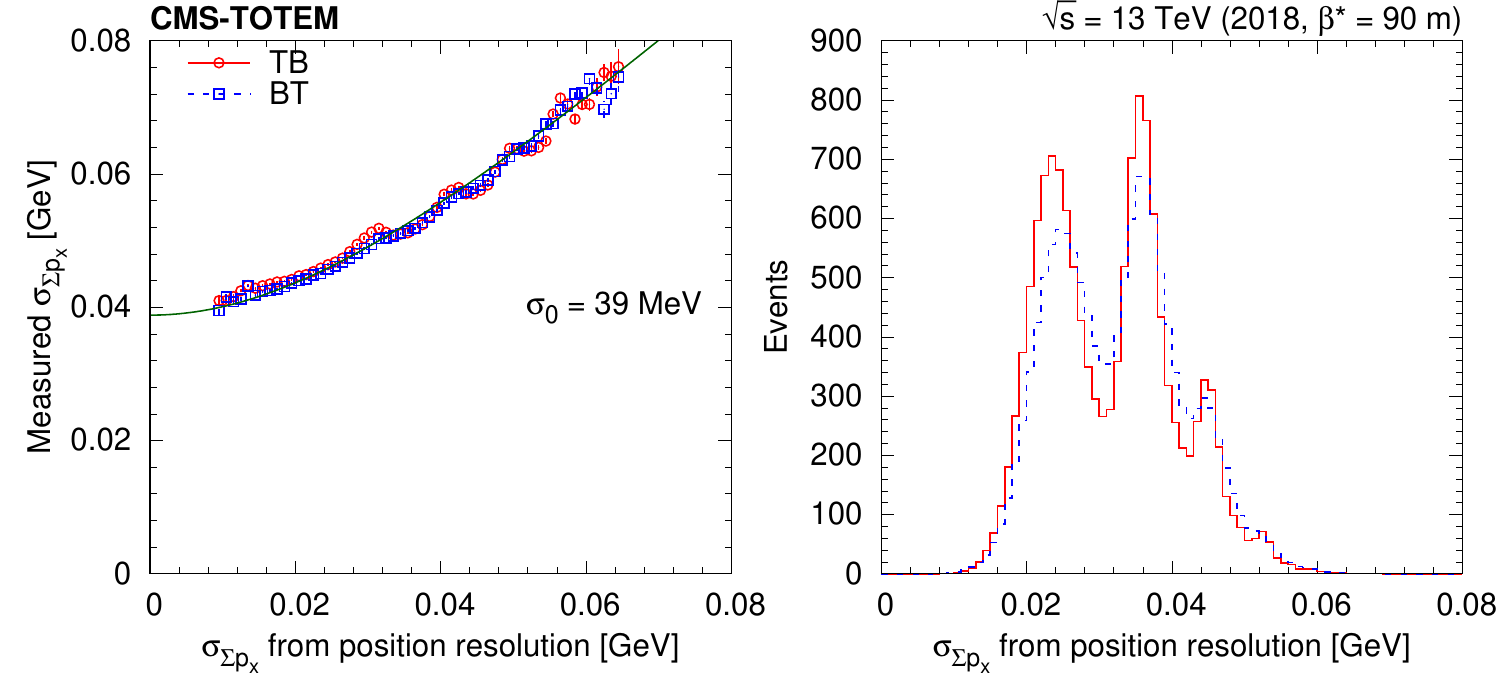}
 \caption{Left: Standard deviation of the momentum sum $\sum p_x$ for the
0-track data (TB and BT configurations), as a function of the predicted
standard deviation from the RP position resolution. Statistical uncertainties
are shown with vertical bars. A fit using the functional form $(\sigma_0^2 +
\sigma_{\sum p_x}^2)^{1/2}$ is plotted with a green curve.  Right: The
occurrence of $\sigma_{\sum p_x}$. The peaks correspond to events with protons
causing one or more two-strip clusters, and hence better resolution.} \label{fig:resolution_x}
\end{figure}

The standard deviation of the momentum sum $\sum p_x$ for the 0-track data as a
function of the predicted standard deviation from the RP position resolution is
shown in Fig.~\ref{fig:resolution_x}, where the result of a fit using the
functional form $(\sigma_0^2 + \sigma_{\sum p_x}^2)^{1/2}$ is indicated. The
constant contribution of $\sigma_0 \approx 39\MeV$ is the result of several
factors: the divergence of both incoming proton beams ($\Delta\sigma_0 \approx
20\MeV$, Section~\ref{sec:lhc_beam}), multiple scattering of the protons within
the RP material, and most importantly the physics process itself. Here
inelastic scattering contributes significantly to $\sigma_0$ if undetected
particles lead to an apparent momentum imbalance.

\clearpage

\begin{figure}
 \centering
 \includegraphics[width=0.49\textwidth]{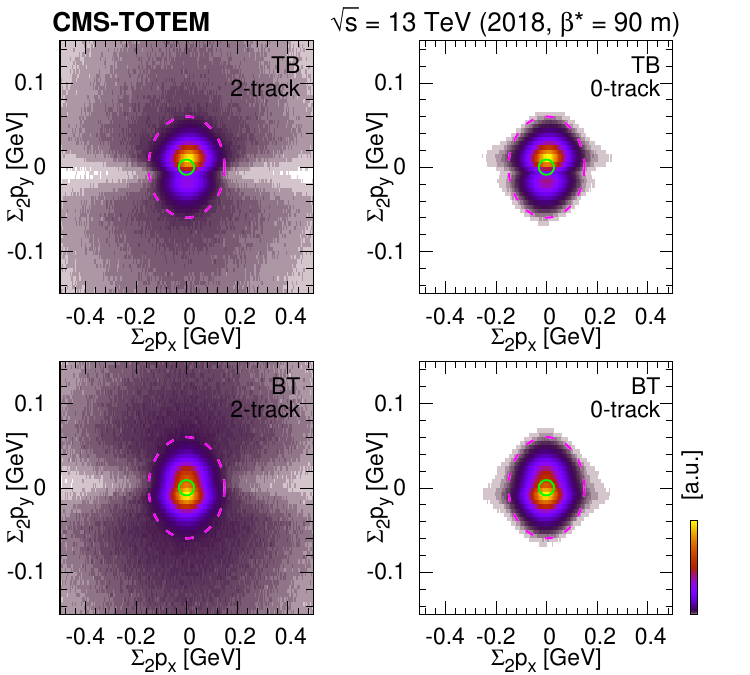}
 \includegraphics[width=0.49\textwidth]{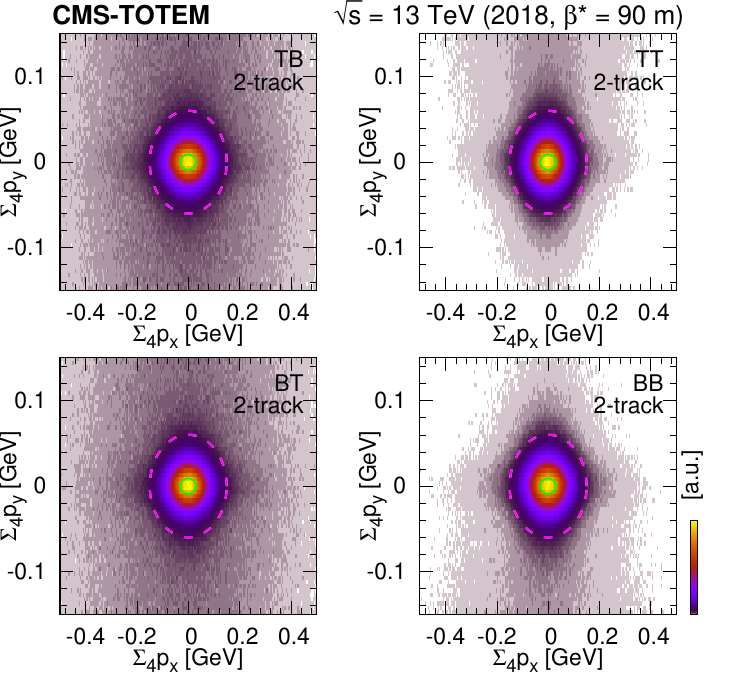}
 \caption{Left: Distribution of the sum of scattered proton momenta $(\sum_2
p_x, \sum_2 p_y)$ for diagonally triggered events (TB and BT). The left column
refers to the 2-track data set, whereas the right one displays the distribution
based on the 0-track data set. Right: Distribution of the sum of scattered
proton and central hadron momenta $(\sum_4 p_x, \sum_4 p_y)$ shown for various
trigger configurations (TB, BT, TT, and BB) for 2-track events. Ellipses with
semi-minor axes of 150\MeV ($x$) and 60\MeV ($y$) are overlaid. The green
circles mark $(0,0)$.} \label{fig:sum2p4p}
\end{figure}

Distributions of the sum of the scattered proton momenta $(\sum_2 p_x, \sum_2
p_y)$, two particles, for diagonally triggered events are shown in
Fig.~\ref{fig:sum2p4p}, left. The measured distributions in the $\sum_2 p_y$
direction are slightly distorted, a consequence of the elastic veto
(Section~\ref{sec:veto}).  Distributions of the sum of scattered proton and
central hadron momenta $(\sum_4 p_x, \sum_4 p_y)$, four particles, shown for
each trigger configuration for 2-track events, are plotted in
Fig.~\ref{fig:sum2p4p}, right. Both distributions are well centred on $(0,0)$.

\begin{figure}[bh]
 \centering
 \includegraphics[width=0.49\textwidth]{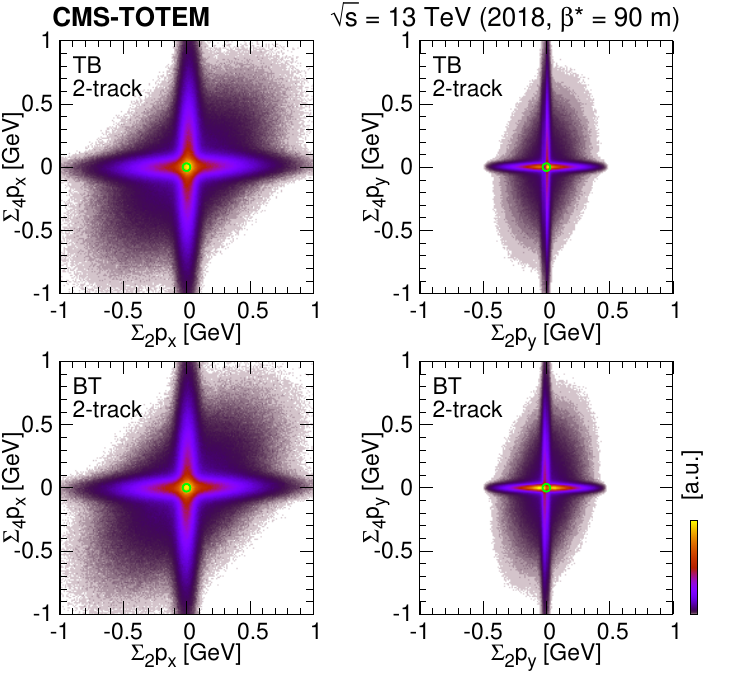}
 \includegraphics[width=0.49\textwidth]{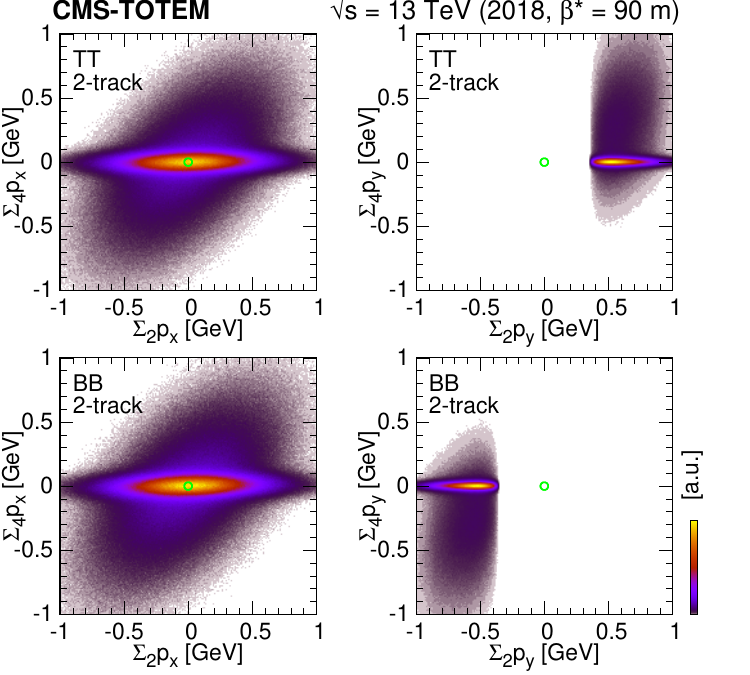}

 \caption{Distribution of the sum of scattered proton and central hadron
momenta and the sum of scattered proton momenta only ($\sum_4 p_x$ and $\sum_2
p_x$, $\sum_4 p_y$ and $\sum_2 p_y$) shown for various trigger configurations
(TB, BT, TT, and BB) for 2-track events. The green circles mark $(0,0)$.}
 \label{fig:sump}
\end{figure}

The distribution of the sum of scattered proton and central hadron momenta and
the sum of scattered proton momenta ($\sum_4 p_x$ and $\sum_2 p_x$, $\sum_4
p_y$ and $\sum_2 p_y$) is shown for each trigger configuration for 2-track
events in Fig.~\ref{fig:sump}. The contributions of true elastic (only two
scattered protons, vertical band) and true central exclusive (two scattered
protons and two central charged hadrons, horizontal band) are well visible. In
addition, a slanted area due to non-exclusive or inelastic background is
present.

The procedures just described are used for the precise transverse momentum
measurement of the scattered protons, for event classification and selection in
the physics analysis of central exclusive production, where two oppositely
charged hadrons are detected by the CMS silicon tracker~\cite{TOTEM:2024dzz}.
 
\clearpage

\section{Summary}
\label{sec:summary}

The Roman pot detectors of the TOTEM experiment are used to reconstruct the
transverse momentum of scattered protons and to estimate the transverse
location of the primary interaction. In this study advanced methods for
track reconstruction, measurements of strip-level detection efficiencies,
cross-checks of beam optics, and the detector alignment procedure are presented,
along with their application in the selection of signal events.

The track reconstruction is performed by finding a common polygonal area
in the intercept-slope plane, thus exploiting all available cluster
information. As a result, an ultimate spatial resolution of $6\text{--}7\mum$
is achieved, which is an order of magnitude smaller than the strip width. The
tool is applied to the relative alignment of detector layers with $\mum$
precision. A tag-and-probe method is used to extract strip-level detection
efficiencies. They are mostly high and constant, but for some strips they
change with time; there are up to 20\% variations. The tracklet
efficiencies are calculated using a probabilistic model, based on the
temporal variation of the hit detection efficiencies. These are functions of
the tracklet location and slope. There are up to 50\% losses in specific but
small areas, to be corrected in the physics analyses.

The alignment of the Roman pot system (8 numbers for each arm) is performed by
means of 16 measured quantities in the horizontal, and 12 in the vertical
direction, resulting in a position accuracy of 3\mum in the horizontal and
60\mum in the vertical directions. The deduced locations of the primary
interaction, the distribution of the scattered proton momenta, and their
correlations confirm the success of the detailed calibration process and
provide a solid ground for exclusive physics analyses based on the
high-$\beta^*$ data set. The developed methods have been successfully applied
in the analysis of central exclusive production events~\cite{TOTEM:2024dzz}.

\begin{acknowledgments}
We congratulate our colleagues in the CERN accelerator departments for the excellent performance of the LHC and thank the technical and administrative staffs at CERN and at other CMS and TOTEM institutes for their contributions to the success of the common CMS-TOTEM effort. In addition, we gratefully acknowledge the computing centres and personnel of the Worldwide LHC Computing Grid and other centres for delivering so effectively the computing infrastructure essential to our analyses. Finally, we acknowledge the enduring support for the construction and operation of the LHC, the CMS and TOTEM detectors, and the supporting computing infrastructure provided by the following funding agencies: SC (Armenia), BMBWF and FWF (Austria); FNRS and FWO (Belgium); CNPq, CAPES, FAPERJ, FAPERGS, and FAPESP (Brazil); MES and BNSF (Bulgaria); CERN; CAS, MoST, and NSFC (China); MINCIENCIAS (Colombia); MSES and CSF (Croatia); RIF (Cyprus); SENESCYT (Ecuador); ERC PRG, RVTT3 and MoER TK202 (Estonia); Academy of Finland,  Magnus Ehrnrooth Foundation, MEC, HIP, and Waldemar von Frenckell Foundation (Finland); CEA and CNRS/IN2P3 (France); SRNSF (Georgia); BMBF, DFG, and HGF (Germany); GSRI (Greece); NKFIH (Hungary); DAE and DST (India); IPM (Iran); SFI (Ireland); INFN (Italy); MSIP and NRF (Republic of Korea); MES (Latvia); LMTLT (Lithuania); MOE and UM (Malaysia); BUAP, CINVESTAV, CONACYT, LNS, SEP, and UASLP-FAI (Mexico); MOS (Montenegro); MBIE (New Zealand); PAEC (Pakistan); MES and NSC (Poland); FCT (Portugal);  MESTD (Serbia); MCIN/AEI and PCTI (Spain); MOSTR (Sri Lanka); Swiss Funding Agencies (Switzerland); MST (Taipei); MHESI and NSTDA (Thailand); TUBITAK and TENMAK (Turkey); NASU (Ukraine); STFC (United Kingdom); DOE and NSF (USA).

\hyphenation{Rachada-pisek} Individuals have received support from the Marie-Curie programme and the European Research Council and Horizon 2020 Grant, contract Nos.\ 675440, 724704, 752730, 758316, 765710, 824093, 101115353, 101002207, and COST Action CA16108 (European Union); the Leventis Foundation; the Alfred P.\ Sloan Foundation; the Alexander von Humboldt Foundation; the Science Committee, project no. 22rl-037 (Armenia); the Belgian Federal Science Policy Office; the Fonds pour la Formation \`a la Recherche dans l'Industrie et dans l'Agriculture (FRIA-Belgium); the F.R.S.-FNRS and FWO (Belgium) under the ``Excellence of Science -- EOS" -- be.h project n.\ 30820817; the Beijing Municipal Science \& Technology Commission, No. Z191100007219010 and Fundamental Research Funds for the Central Universities (China); the Ministry of Education, Youth and Sports (MEYS) of the Czech Republic; the Shota Rustaveli National Science Foundation, grant FR-22-985 (Georgia); the Deutsche Forschungsgemeinschaft (DFG), among others, under Germany's Excellence Strategy -- EXC 2121 ``Quantum Universe" -- 390833306, and under project number 400140256 - GRK2497; the Hellenic Foundation for Research and Innovation (HFRI), Project Number 2288 (Greece); the Hungarian Academy of Sciences, the New National Excellence Program - \'UNKP, the NKFIH research grants K 131991, K 133046, K 138136, K 143460, K 143477, K 146913, K 146914, K 147048, 2020-2.2.1-ED-2021-00181, TKP2021-NKTA-64, and 2021-4.1.2-NEMZ\_KI-2024-00036 (Hungary); the Council of Science and Industrial Research, India; ICSC -- National Research Centre for High Performance Computing, Big Data and Quantum Computing and FAIR -- Future Artificial Intelligence Research, funded by the NextGenerationEU program (Italy); the Latvian Council of Science; the Ministry of Education and Science, project no. 2022/WK/14, and the National Science Centre, contracts Opus 2021/41/B/ST2/01369 and 2021/43/B/ST2/01552 (Poland); the Funda\c{c}\~ao para a Ci\^encia e a Tecnologia, grant CEECIND/01334/2018 (Portugal); the National Priorities Research Program by Qatar National Research Fund; MCIN/AEI/10.13039/501100011033, ERDF ``a way of making Europe", and the Programa Estatal de Fomento de la Investigaci{\'o}n Cient{\'i}fica y T{\'e}cnica de Excelencia Mar\'{\i}a de Maeztu, grant MDM-2017-0765 and Programa Severo Ochoa del Principado de Asturias (Spain); the Chulalongkorn Academic into Its 2nd Century Project Advancement Project, and the National Science, Research and Innovation Fund via the Program Management Unit for Human Resources \& Institutional Development, Research and Innovation, grant B39G670016 (Thailand); the Kavli Foundation; the Nvidia Corporation; the SuperMicro Corporation; the Welch Foundation, contract C-1845; and the Weston Havens Foundation (USA).
\end{acknowledgments}

\bibliography{auto_generated}
\cleardoublepage \appendix\section{The CMS Collaboration \label{app:collab}}\begin{sloppypar}\hyphenpenalty=5000\widowpenalty=500\clubpenalty=5000\cmsinstitute{Yerevan Physics Institute, Yerevan, Armenia}
{\tolerance=6000
A.~Hayrapetyan, A.~Tumasyan\cmsAuthorMark{1}\cmsorcid{0009-0000-0684-6742}
\par}
\cmsinstitute{Institut f\"{u}r Hochenergiephysik, Vienna, Austria}
{\tolerance=6000
W.~Adam\cmsorcid{0000-0001-9099-4341}, J.W.~Andrejkovic, L.~Benato\cmsorcid{0000-0001-5135-7489}, T.~Bergauer\cmsorcid{0000-0002-5786-0293}, S.~Chatterjee\cmsorcid{0000-0003-2660-0349}, K.~Damanakis\cmsorcid{0000-0001-5389-2872}, M.~Dragicevic\cmsorcid{0000-0003-1967-6783}, P.S.~Hussain\cmsorcid{0000-0002-4825-5278}, M.~Jeitler\cmsAuthorMark{2}\cmsorcid{0000-0002-5141-9560}, N.~Krammer\cmsorcid{0000-0002-0548-0985}, A.~Li\cmsorcid{0000-0002-4547-116X}, D.~Liko\cmsorcid{0000-0002-3380-473X}, I.~Mikulec\cmsorcid{0000-0003-0385-2746}, J.~Schieck\cmsAuthorMark{2}\cmsorcid{0000-0002-1058-8093}, R.~Sch\"{o}fbeck\cmsAuthorMark{2}\cmsorcid{0000-0002-2332-8784}, D.~Schwarz\cmsorcid{0000-0002-3821-7331}, M.~Sonawane\cmsorcid{0000-0003-0510-7010}, W.~Waltenberger\cmsorcid{0000-0002-6215-7228}, C.-E.~Wulz\cmsAuthorMark{2}\cmsorcid{0000-0001-9226-5812}
% ...
\par}
\vskip\cmsinstskip
\dag:~Deceased\\
$^{1}$Also at Yerevan State University, Yerevan, Armenia\\
$^{2}$Also at TU Wien, Vienna, Austria\\
% ...
\section{The TOTEM Collaboration\label{app:totem}}
\newcommand{\AddAuthor}[2]{#1$^{#2}$,\ }
\newcommand\AddAuthorLast[2]{#1$^{#2}$}
\noindent
\AddAuthor{G.~Antchev\cmsorcid{0000-0003-3210-5037}}{a}
\AddAuthor{P.~Aspell}{8}
\AddAuthor{I.~Atanassov\cmsorcid{0000-0002-5728-9103}}{a}
\AddAuthor{V.~Avati}{7,8}
\AddAuthor{J.~Baechler}{8}
\AddAuthor{C.~Baldenegro~Barrera\cmsorcid{0000-0002-6033-8885}}{11}
\AddAuthor{V.~Berardi\cmsorcid{0000-0002-8387-4568}}{4a,4b}
\AddAuthor{M.~Berretti\cmsorcid{0000-0003-4122-8282}}{2a}
\AddAuthor{V.~Borshch\cmsorcid{0000-0002-5479-1982}}{12}
\AddAuthor{E.~Bossini\cmsorcid{0000-0002-2303-2588}}{6a}
\AddAuthor{U.~Bottigli\cmsorcid{0000-0002-0666-3433}}{6b}
\AddAuthor{M.~Bozzo\cmsorcid{0000-0002-1715-0457}}{5a,5b}
\AddAuthor{H.~Burkhardt}{8}
\AddAuthor{F.S.~Cafagna\cmsorcid{0000-0002-7450-4784}}{4a}
\AddAuthor{M.G.~Catanesi\cmsorcid{0000-0002-2987-7688}}{4a}
\AddAuthor{M.~Deile\cmsorcid{0000-0001-5085-7270}}{8}
\AddAuthor{F.~De~Leonardis}{4a,4c}
\AddAuthor{M.~Doubek\cmsorcid{0000-0002-0026-9558}}{1c}
\AddAuthor{D.~Druzhkin}{3a,8}
\AddAuthor{K.~Eggert}{10}
\AddAuthor{V.~Eremin}{c}
\AddAuthor{A.~Fiergolski}{8}
\AddAuthor{L.~Forthomme\cmsorcid{0000-0002-3302-336X}}{2a,2b}
\AddAuthor{F.~Garcia\cmsorcid{0000-0002-4023-7964}}{2a}
\AddAuthor{V.~Georgiev}{1a}
\AddAuthor{S.~Giani}{8}
\AddAuthor{L.~Grzanka\cmsorcid{0000-0002-3599-854X}}{7}
\AddAuthor{J.~Hammerbauer}{1a}
\AddAuthor{T.~Isidori\cmsorcid{0000-0002-7934-4038}}{11}
\AddAuthor{V.~Ivanchenko\cmsorcid{0000-0002-1844-5433}}{12}
\AddAuthor{M.~Janda\cmsorcid{0000-0001-5736-6183}}{1c}
\AddAuthor{A.~Karev}{8}
\AddAuthor{J.~Ka\v{s}par\cmsorcid{0000-0001-5639-2267}}{1b,8}
\AddAuthor{B.~Kaynak\cmsorcid{0000-0003-3857-2496}}{9}
\AddAuthor{J.~Kopal\cmsorcid{0000-0001-9751-7409}}{8}
\AddAuthor{V.~Kundr\'{a}t\cmsorcid{0000-0003-2868-5550}}{1b}
\AddAuthor{S.~Lami\cmsorcid{0000-0001-9492-0147}}{6a}
\AddAuthor{R.~Linhart}{1a}
\AddAuthor{C.~Lindsey}{11}
\AddAuthor{M.V.~Lokaj\'{i}\v{c}ek\cmsorcid{0000-0002-2052-1220}$^{\textrm{\dag,}}$}{1b}
\AddAuthor{L.~Losurdo\cmsorcid{0000-0002-4964-7951}}{6b}
\AddAuthor{F.~Lucas~Rodr\'{i}guez}{8}
\AddAuthor{M.~Macr\'{i}}{\dag,5a}
\AddAuthor{M.~Malawski\cmsorcid{0000-0001-6005-0243}}{7}
\AddAuthor{N.~Minafra\cmsorcid{0000-0003-4002-1888}}{11}
\AddAuthor{S.~Minutoli}{5a}
\AddAuthor{K.~Misan\cmsorcid{0009-0007-2416-042X}}{7}
\AddAuthor{T.~Naaranoja\cmsorcid{0000-0001-5797-7929}}{2a,2b}
\AddAuthor{F.~Nemes\cmsorcid{0000-0002-1451-6484}}{3a,3b,8}
\AddAuthor{H.~Niewiadomski}{10}
\AddAuthor{E.~Oliveri\cmsorcid{0000-0002-0832-6975}}{8}
\AddAuthor{F.~Oljemark\cmsorcid{0000-0003-0121-2761}}{2a,2b}
\AddAuthor{M.~Oriunno}{b}
\AddAuthor{K.~\"{O}sterberg\cmsorcid{0000-0003-4807-0414}}{2a,2b}
\AddAuthor{S.~Ozkorucuklu\cmsorcid{0000-0001-5153-9266}}{9}
\AddAuthor{P.~Palazzi\cmsorcid{0000-0002-4861-391X}}{8}
\AddAuthor{V.~Passaro\cmsorcid{0000-0003-0802-4464}}{4a,4c}
\AddAuthor{Z.~Peroutka}{1a}
\AddAuthor{O.~Potok\cmsorcid{0009-0005-1141-6401}}{9}
\AddAuthor{J.~Proch\'{a}zka\cmsorcid{0000-0002-2774-2245}}{1b,\dag}
\AddAuthor{M.~Quinto\cmsorcid{0000-0002-6363-6132}}{4a,4b}
\AddAuthor{E.~Radermacher}{8}
\AddAuthor{E.~Radicioni\cmsorcid{0000-0002-2231-1067}}{4a}
\AddAuthor{F.~Ravotti\cmsorcid{0000-0002-5709-6934}}{8}
\AddAuthor{C.~Royon\cmsorcid{0000-0002-7672-9709}}{11}
\AddAuthor{G.~Ruggiero}{8}
\AddAuthor{H.~Saarikko}{2a,2b}
\AddAuthor{V.D.~Samoylenko}{c}
\AddAuthor{A.~Scribano\cmsorcid{0000-0002-4338-6332}}{6a,11}
\AddAuthor{J.~\v{S}irok\'{y}}{1a}
\AddAuthor{J.~Smajek}{8}
\AddAuthor{W.~Snoeys\cmsorcid{0000-0003-3541-9066}}{8}
\AddAuthor{R.~Stefanovitch}{3a,8}
\AddAuthor{C.~Taylor\cmsorcid{0000-0001-6816-8051}}{10}
\AddAuthor{E.~Tcherniaev\cmsorcid{0000-0002-3685-0635}}{12}
\AddAuthor{N.~Turini\cmsorcid{0000-0002-9395-5230}}{6b}
\AddAuthor{O.~Urban}{1a}
\AddAuthor{V.~Vacek\cmsorcid{0000-0001-9584-0392}}{1c}
\AddAuthor{O.~Vavroch}{1a}
\AddAuthor{J.~Welti}{2a,2b}
\AddAuthor{J.~Williams\cmsorcid{0000-0002-9810-7097}}{11}
\AddAuthorLast{J.~Zich}{1a}
\vskip 4pt plus 4pt
\let\thefootnote\relax
\newcommand{\AddInstitute}[2]{${}^{#1}$#2\\}
\newcommand{\AddExternalInstitute}[2]{\footnote{${}^{#1}$ #2}}
\noindent
\dag Deceased\\
\AddInstitute{1a}{University of West Bohemia, Pilsen, Czech Republic}
\AddInstitute{1b}{Institute of Physics of the Academy of Sciences of the Czech Republic, Prague, Czech Republic}
\AddInstitute{1c}{Czech Technical University, Prague, Czech Republic}
\AddInstitute{2a}{Helsinki Institute of Physics, University of Helsinki, Helsinki, Finland}
\AddInstitute{2b}{Department of Physics, University of Helsinki, Helsinki, Finland}
\AddInstitute{3a}{Wigner Research Centre for Physics, RMKI, Budapest, Hungary}
\AddInstitute{3b}{MATE Institute of Technology KRC, Gy\"{o}ngy\"{o}s, Hungary}
\AddInstitute{4a}{INFN Sezione di Bari, Bari, Italy}
\AddInstitute{4b}{Dipartimento Interateneo di Fisica di Bari, University of Bari, Bari, Italy}
\AddInstitute{4c}{Dipartimento di Ingegneria Elettrica e dell'Informazione --- Politecnico di Bari, Bari, Italy}
\AddInstitute{5a}{INFN Sezione di Genova, Genova, Italy}
\AddInstitute{5b}{Universit\`{a} degli Studi di Genova, Genova, Italy}
\AddInstitute{6a}{INFN Sezione di Pisa, Pisa, Italy}
\AddInstitute{6b}{Universit\`{a} degli Studi di Siena and Gruppo Collegato INFN di Siena, Siena, Italy}
\AddInstitute{7}{Akademia G\'{o}rniczo-Hutnicza (AGH) University of Krak\'ow, Faculty of Computer Science, Krak\'ow, Poland}
\AddInstitute{8}{CERN, Geneva, Switzerland}
\AddInstitute{9}{Istanbul University, Istanbul, Turkey}
\AddInstitute{10}{Case Western Reserve University, Department of Physics, Cleveland, Ohio, USA}
\AddInstitute{11}{The University of Kansas, Lawrence, Kansas, USA}
\AddInstitute{12}{Authors affiliated with an institute or an international laboratory covered by a cooperation agreement with CERN}
\AddExternalInstitute{a}{INRNE-BAS, Institute for Nuclear Research and Nuclear Energy, Bulgarian Academy of Sciences, Sofia, Bulgaria}
\AddExternalInstitute{b}{SLAC, Stanford University, California, USA}
\AddExternalInstitute{c}{Authors affiliated with an institute or an international laboratory covered by a cooperation agreement with CERN}
\end{sloppypar}
\end{document}